\documentclass[11pt]{article}
\usepackage[utf8]{inputenc}
\usepackage[english]{babel}
\usepackage[active]{srcltx}
\usepackage{hyperref}
\usepackage[all]{xy}
\usepackage{amsfonts}
\usepackage{amsthm}
\usepackage{enumerate}
\usepackage{mathrsfs}
\usepackage{multicol}
\usepackage[all]{xy}
\usepackage{amsfonts}
\usepackage{latexsym,amsmath,amssymb}
\usepackage[usenames,dvipsnames]{color}
\usepackage{xfrac}
\usepackage{pstricks}
\usepackage[absolute,overlay]{textpos}
\usepackage{graphics,wrapfig,times}

\usepackage{color}
\definecolor{DarkBlue}{rgb}{0.1,0.1,0.5}
\definecolor{Red}{rgb}{0.9,0.1,0.1}
\definecolor{Green}{rgb}{0.3,0.7,0.0}
\definecolor{green2}{rgb}{0.1,0.7,0.2}
\definecolor{blue2}{rgb}{0.0,0.6,0.7}
\definecolor{pink}{rgb}{1,0.0,1}
\definecolor{orange}{rgb}{0.9,0.0,0.1}

\newtheorem{theo}{Theorem}

\newtheorem{prop}{Proposition}

\newtheorem{definition}{Definition}
\newtheorem{remark}{Remark}

\renewcommand{\d}{\mathrm{d}}
\newcommand{\derpar}[2]{\displaystyle\frac{\partial{#1}}{\partial{#2}}}
\newcommand{\derpars}[3]{\displaystyle\frac{\partial^2{#1}}{\partial{#2}\partial{#3}}}

\newcommand{\Lag}{\mathcal{L}}

\newcommand{\vf}{\mathfrak{X}}
\newcommand{\df}{\Omega}

\newcommand{\Tan}{\mathrm{T}}
\newcommand{\inn}{{\mathop{i}\nolimits}}
\newcommand{\Lie}{\mathop{\mathrm{L}}\nolimits}

\newcommand{\bal}{\begin{align*}}
\newcommand{\eal}{\end{align*}}
\def\beq{\begin{equation}}
\def\eeq{\end{equation}}
\def\bea{\begin{eqnarray}}
\def\eea{\end{eqnarray}}
\def\beann{\begin{eqnarray*}}
\def\eeann{\end{eqnarray*}}
\def\ben{\begin{enumerate}}
\def\een{\end{enumerate}}
\def\bit{\begin{itemize}}
\def\eit{\end{itemize}}

\def\dst{\displaystyle}

\def\vf{\mathfrak X}
\def\df{{\mit\Omega}}
\def\Lag{{\cal L}}

\def\d{{\rm d}}

\def\Tan{{\rm T}}
\def\Lie{\mathop{\rm L}\nolimits}
\def\inn{\mathop{i}\nolimits}
\def\Cinfty{{\rm C}^\infty}

\title{\sc Symmetries and gauge symmetries in multisymplectic first and
second-order Lagrangian field theories:
electromagnetic and gravitational fields}
\author{
$^1${\sc  Jordi Gaset \thanks{{\bf e}-{\it mail}:
   jordi.gaset@uab.cat. (ORCID: 0000-0001-8796-3149).} },
   $^2${\sc Narciso Rom\'an-Roy \thanks{{\bf e}-{\it mail}:
   narciso.roman@upc.edu. (ORCID: 0000-0003-3663-9861).} }
\\[1ex]
\normalsize\itshape\sffamily 
$^1$Department of Physics,
Universitat Aut\`onoma de Barcelona,
Bellaterra, Spain.
\\[1ex]
\normalsize\itshape\sffamily 
$^2$Department of Mathematics,
Universitat Polit\`ecnica de Catalunya,
Barcelona, Spain.
}

\begin{document}

\maketitle

\pagestyle{myheadings}
\markright{\rm J. Gaset, N. Rom\'an-Roy:
   \sl Symmetries in Lagrangian field theories: Electromagnetism and Gravitation.}
\maketitle
\thispagestyle{empty}

\begin{abstract}
Symmetries and, in particular,
Cartan (Noether) symmetries and conserved quantities (conservation laws) are studied
for the multisymplectic formulation of first and second order Lagrangian classical field theories.
Noether-type theorems are stated in this geometric framework.
The concept of gauge symmetry and its geometrical meaning are also discussed
in this formulation.
The results are applied to study Noether and gauge symmetries for the
multisymplectic description of the electromagnetic and the gravitational theory; in particular, the Einstein--Hilbert
and the Einstein--Palatini approaches.
\end{abstract}


\noindent\textbf{MSC\,2020 codes:}
{\it Primary}: 53D42, 70S05, 83C05; {\it Secondary}: 35Q75, 35Q76, 53Z05, 70H50,  83C99.


\medskip
\setcounter{tocdepth}{2}
{
\small
\def\addvspace#1{\vskip 1pt}
\parskip 0pt plus 0.1mm
\tableofcontents
}

\section{Introduction}
\label{intro}

It is well known that symmetries have enormous relevance in physical theories
and, in general, in the treatment and resolution of differential equations
modelling them. 
This is due to the fact that the presence of symmetries leads 
to the existence of conservation laws or conserved quantities
which, in addition to helping the integration of these equations, 
highlight fundamental properties of physical systems.
In this sense, the work of {\it E. Noether} 
at the beginning of the 20th century provides fundamental results 
on this topic \cite{KS-2011}.

From a geometric perspective, symmetries of mechanical systems and
classical field theories are usually stated by demanding the
invariance of the underlying geometric structures and/or the
dynamical elements which characterize these systems
(for a review in the case of mechanics see, for instance, 
\cite{RR-2020} and the references cited therein).
In the case of the {\sl multisymplectic description}
of field theories, these are the {\sl multisymplectic forms}
which are defined in the jet bundles and the multimomentum bundles
where the theory is developed.
These forms are constructed from the Lagrangian
which describes the system, using the canonical elements
which these bundles are endowed with
\cite{art:Aldaya_Azcarraga78_2,EMR-96,Gc-73,GS-73,HK-04,
KrupkaStepanova,art:Munoz85,book:Saunders89}.
The study of symmetries and their associated conservation laws
in this framework has been carried out in many papers
(see, for instance, \cite{art:deLeon_etal2004,EMR-99b,FS-2012,art:GPR-2016,RWZ-2016} and the references therein).

One of the main characteristics of certain kinds of physical theories
is the so-called {\sl gauge invariance}, which is a consequence
of the existence of
a particular type of symmetry called {\sl gauge symmetry}.
This is a property which is associated with physical systems described by {\sl singular Lagrangians}.
Gauge symmetries have their own geometric characterization
which is related to the fact that the multisymplectic forms 
constructed from the (singular) Lagrangians are degenerated 
(and then they are called {\sl premultisymplectic forms\/}).

In this review paper, our aim is to present an accurate geometric description
of symmetries in classical field theories of first and second-order type.
First, we introduce the standard classical symmetries 
{\sl of Noether type} and state the {\sl Noether theorem}
which gives the way of obtaining the corresponding 
associated conservation laws.
Second, we discuss in detail the geometrical meaning and the
characteristics of gauge symmetries;
clarifying several geometric aspects
that are not usually analysed in most treaties on this subject.
Finally, we apply the results to the case of the two fundamental classical theories: 
Electromagnetism (Maxwell theory) and Gravitation (General Relativity);
in this last case, considering the two more  basic models;
i.e., the {\sl Einstein--Hilbert }and the {\sl Einstein--Palatini} approaches (the first and third are first-order field theories,
but the second is second-order).

The organization of the paper is as follows:
Section \ref{previous} is devoted to summarize the multisymplectic
Lagrangian  formulation of first and second-order classical field theories.
In Section \ref{S3} we study symmetries, conservation laws
and gauge symmetries in this multisymplectic setting.
The analysis of symmetries of Electromagnetism is analyzed in Section \ref{EM}
and, finally, in Section \ref{S4} we study Noether and gauge symmetries
of both models in General Relativity.

All the manifolds are real, second countable and $\Cinfty$. The maps and the structures are $\Cinfty$.  Sum over repeated indices is understood.
Along this paper, we use the notation of multi-indices:
a multi-index $I$ is an element of $\mathbb{Z}^m$ where every
component is positive, the $i$th position of the multi-index is denoted $I(i)$, 
and $\displaystyle |I| = \sum_{i=1}^{m} I(i)$ is the length of the multi-index.
The equality $|I| = k$ means that the expression is taken for
every multi-index of length $k$.
Furthermore, the element $1_i\in\mathbb{Z}^m$ is defined as $1_i(j)=\delta_i^j$.
Finally, $n(ij)$ is a combinatorial factor with $n(ij)=1$ for $i=j$, and $n(ij)=2$ for $i\neq j$.


\section{Lagrangian field theories in jet bundles}
\label{previous}

\subsection{Higher-order jet bundles. Multivector fields in jet bundles}
\label{S2}

(See \cite{art:Echeverria_Munoz_Roman98,IEMR-2012,book:Saunders89} for details).

Let $E \stackrel{\pi}{\longrightarrow} M$ be a fiber bundle over
an orientable $m$-dimensional manifold $M$, with $\dim E = m + n$.
The {\sl \textbf{$k$th-order jet bundle}} of the projection $\pi$
is the manifold of the $k$-jets (equivalence classes) of local sections of $\pi$, 
$\phi \in \Gamma(\pi)$,
and is denoted $J^k\pi$.
Points in $J^k\pi$ are denoted by $j^k_x\phi$, 
with $x \in M$ and $\phi \in \Gamma(\pi)$ being a representative
of the equivalence class.
$J^k\pi$ is endowed with the following natural projections: if $1\leqslant r \leqslant k$,
$$
\begin{array}{rcl}
\pi^k_r \colon J^k\pi & \longrightarrow & J^r\pi \\
j^k_x\phi & \longmapsto & j^r_x\phi
\end{array}
\quad \ \quad
\begin{array}{rcl}
\pi^k \colon J^k\pi & \longrightarrow & E \\
j^k_x\phi & \longmapsto & \phi(x)
\end{array}
\quad \ \quad
\begin{array}{rcl}
\bar{\pi}^k \colon J^k\pi & \longrightarrow & M \\
j^k_x\phi & \longmapsto & x
\end{array}
$$
where $\pi^s_r\circ\pi^k_s = \pi^k_r$, $\pi^k_0 = \pi^k$, $\pi^k_k = \textnormal{Id}_{J^k\pi}$,
and $\bar{\pi}^k = \pi \circ \pi^k$.
We denote $\omega$ the volume form in $M$ 
and all its pull-backs to every $J^r\pi$.
If $(x^i,y^\alpha)$, $1 \leqslant i \leqslant m$, $1 \leqslant \alpha \leqslant n$, are local coordinates in $E$ adapted to the bundle structure,
such that $\omega=\d x^1\wedge\ldots\wedge\d x^m\equiv\d^mx$;
then local coordinates in $J^k\pi$ are
denoted $(x^i,y_I^\alpha)$, with $0 \leqslant |I| \leqslant k$.

If $\phi \in \Gamma(\pi)$, the {\sl \textbf{$k$th prolongation}} of $\phi$ to $J^k\pi$ is denoted $j^k\phi \in \Gamma(\bar{\pi}^k)$.
Then, a section $\psi \in \Gamma(\bar{\pi}^{k})$ is {\sl \textbf{holonomic}} 
if $j^k(\pi^{k} \circ \psi) = \psi$; that is, 
$\psi$ is the $k$th prolongation of the section $\phi=\pi^{k}\circ\psi\in\Gamma(\pi)$.

In an analogous way, let $\Phi\colon E\to E$ be a diffeomorphism and
and $\Phi_M\colon M\to M$ the diffeomorphism
induced on the basis;  then the {\sl \textbf{canonical lift}} 
of $\Phi$ to $J^k\pi$ is the map 
$j^k\Phi\colon J^k\pi\longrightarrow J^k\pi$ defined by
$$
(j^k\Phi)(j^k_x\phi):=j^k(\Phi\circ\phi\circ\Phi_M^{-1})(\Phi_M(x))
\quad ; \quad \mbox{\rm for $j^k_x\phi\in J^k\pi$} \ .
$$
Then, if $Y\in\vf(E)$, the {\sl \textbf {canonical lift}} of $Y$
to $J^k\pi$ is the vector field $j^1Y\in\vf(J^k\pi)$ whose associated
local one-parameter groups of diffeomorphisms are
the canonical lifts of the local
one-parameter groups of diffeomorphisms of $Y$.
In coordinates, if 
$\dst Y=f^i\derpar{}{x^i}+g^\alpha\derpar{}{y^\alpha}$; 
for instance, for $k=1$ we have that
\beann
j^1Y&=& f^i\derpar{}{x^i}+g^\alpha\derpar{}{y^\alpha}+
\left(\derpar{g^\alpha}{x^i}-
y^\alpha_j\left(\derpar{f^j}{x^i}+
y^\beta_i\derpar{f^j}{y^\beta}\right)+y^\beta_i\derpar{g^\alpha}{y^\beta}\right)
\derpar{}{y^\alpha_i} \ .
\eeann

Another special kind of vector fields are the {\sl \textbf{coordinate total derivatives}}
\cite{pere,book:Saunders89}:
$$
D_{i}=\frac{\partial}{\partial x^i}+\sum_{|I|=0}^k y_{I+1_i}^\alpha\frac{\partial}{\partial y_{I}^\alpha}\ ,
$$
and for $f\in\Cinfty(J^k\pi)$, we have that $D_i f=\Lie(D_i) f$.

An {\sl \textbf{$m$-multivector field}} in $J^k\pi$ is a skew-symmetric contravariant 
tensor field of order $m$ in $J^k\pi$. The set of $m$-multivector fields 
in $J^k\pi$ is denoted $\vf^m (J^k\pi)$.
A multivector field $\mathbf{X}\in\vf^m(J^k\pi)$ is {\sl \textbf{locally decomposable}} if,
for every $j^k_x\phi\in J^k\pi$, there is an open neighbourhood  $U\subset J^k\pi$, 
with $j^k_x\phi\in U$,
and $X_1,\ldots ,X_m\in\vf (U)$ such that $\mathbf{X}\vert_U=X_1\wedge\ldots\wedge X_m$.
Locally decomposable $m$-multivector fields are locally associated with $m$-dimensional
distributions $D\subset\Tan J^k\pi$, and multivector fields associated with
the same distribution make an {\sl equivalence class} $\{ {\bf X}\}$ in the set $\vf^m(J^k\pi)$.
Then,
$\mathbf{X}$ is {\sl \textbf{integrable}} if its associated distribution is integrable. 
In particular,
$\mathbf{X}$ is {\sl \textbf{holonomic}} if
it is integrable and  its integral sections are holonomic sections of $\bar\pi^k$.

A multivector field $\mathbf{X}\in\mathfrak{X}^m(J^k\pi)$ 
is \emph{$\bar\pi^k$-transverse} if, at every point $j^k_x\phi\in J^k\pi$,
we have that $(\inn(\mathbf{X})(\bar\pi^{k*}\beta))_{j^k_x\phi}\not= 0$; 
for every $\beta\in\Omega^m(M)$ such that
$\beta_{\bar\pi^k(j^k_x\phi)}\not= 0$. 
If $\mathbf{X}\in\mathfrak{X}^m(J^k\pi)$ is
integrable, then it is       
$\bar\pi^k$-transverse if, and only if, its integral manifolds are local sections of
$\bar\pi^k\colon J^k\pi\to M$.
In this case, if $\psi\colon U\subset M\to J^k\pi$ is a local
section with $\psi (x)=j^k_x\phi$ and $\psi (U)$ is the integral manifold 
of $\mathbf{X}$ at $j^k_x\phi$; then  $T_{j^k_x\phi}({\rm Im}\,\psi) = \mathcal{D}_{j^k_x\phi}(\mathbf{X})$
and $\psi$ is an integral section of ${\bf X}$.
(See \cite{EMR-99b} for more details).

For every $\mathbf{X}\in\mathfrak{X}^m(J^k\pi)$, 
there exist $X_1,\ldots ,X_r\in\mathfrak{X} (U)$ such that
$$
\mathbf{X}\vert_{U}=\sum_{1\leq i_1<\ldots <i_m\leq r} f^{i_1\ldots i_m}X_{i_1}\wedge\ldots\wedge X_{i_m} \, ,
$$
with $f^{i_1\ldots i_m} \in C^\infty (U)$, $m \leqslant r\leqslant{\rm dim}\,J^k\pi$.
Therefore, the condition of ${\bf X}$ to be integrable is locally equivalent to
$[X_i,X_j]=0$, for $i,j=1,\ldots,m$.
If ${\bf X},{\bf X}'\in\{ {\bf X}\}$ then, for every $U\subset J^k\pi$,
there exists a non-vanishing function $f\in\Cinfty(U)$ such that 
${\bf X}'=f{\bf X}$ on $U$.

In natural coordinates, a locally decomposable and 
$\pi^k$--transverse multivector field ${\bf X}\in\vf^m(J^k\pi)$
can be written as
$$
{\bf X}=f\bigwedge_{i=1}^m\left(\frac{\partial}{\partial x^i}+
F^\alpha_i\frac{\partial}{\partial y^\alpha}+
F^\alpha_{I,i}\frac{\partial}{\partial y_{I}^\alpha}\right)
\quad , \quad (1\leq\vert I\vert\leq k)\ ,
$$
and, if it is holonomic,
\beq
{\bf X}=f\bigwedge_{i=1}^m\left(\frac{\partial}{\partial x^i}+
y^\alpha_i\frac{\partial}{\partial y^\alpha}+
\sum_{|I|=1}^{k-1} y_{I+1_i}^\alpha\frac{\partial}{\partial y_{I}^\alpha}+
F^\alpha_{K,i}\frac{\partial}{\partial y_{K}^\alpha}\right)
\quad , \quad (\vert K\vert=k)\ .
\label{locholmvf}
\eeq

If $\Omega\in\df^p(J^k\pi)$ and $\mathbf{X}\in\mathfrak{X}^m(J^k\pi)$,
the {\sl \textbf{contraction}} between ${\bf X}$ and $\Omega$ is
the natural contraction between tensor fields; in particular,
it gives zero when $p<m$ and, if $p\geq m$,
$$
 \inn({\bf X})\Omega\mid_{U}:= \sum_{1\leq i_1<\ldots <i_m\leq
 r}f^{i_1\ldots i_m} \inn(X_1\wedge\ldots\wedge X_m)\Omega 
=
 \sum_{1\leq i_1<\ldots <i_m\leq r}f^{i_1\ldots i_m} \inn
 (X_1)\ldots\inn (X_m)\Omega \ .
$$
The {\sl \textbf{Lie derivative}} of $\Omega$ with respect to ${\bf X}$ is defined as the graded bracket (of degree $m-1$)
 $$
\Lie({\bf X})\Omega:=[\d , \inn ({\bf X})]\Omega=
(\d\inn ({\bf X})-(-1)^m\inn ({\bf X})\d)\Omega \ .
 $$
If ${\bf X}\in\vf^i({\cal M})$ and ${\bf Y}\in\vf^j(J^k\pi)$,
the {\sl \textbf{Schouten-Nijenhuis bracket}} of ${\bf X},{\bf Y}$
is the bilinear assignment ${\bf X},{\bf Y}\mapsto [{\bf X},{\bf Y}]$, 
where $[{\bf X},{\bf Y}]$ is a $(i+j-1)$-multivector field obtained 
as the graded commutator of $\Lie ({\bf X})$ and $\Lie ({\bf Y})$
(which is an operation of degree $i+j-2$),
 $$
\Lie([{\bf X},{\bf Y}]):=[\Lie ({\bf X}), \Lie ({\bf Y})] \ .
 $$
It is also called the {\sl \textbf{Lie derivative}} of ${\bf Y}$ with respect to ${\bf X}$,
and is denoted as $\Lie({\bf X}){\bf Y}:=[{\bf X},{\bf Y}]$.

\subsection{First and second-order Lagrangian field theories}

(See \cite{art:Aldaya_Azcarraga78_2,EMR-96,art:Echeverria_Munoz_Roman98,Gc-73,GS-73,pere,book:Saunders89} for details).

Let  $\pi \colon E \to M$ be the {\sl configuration bundle} of a
{\sl first} or {\sl second order} classical field theory.

For first-order field theories,
we have a first-order Lagrangian density $\Lag \in \df^{m}(J^1\pi)$, which is a $\overline{\pi}^1$-semibasic m-form and then
$\mathcal{L}=L\,(\overline{\pi}^1)^*\eta$, 
where $L\in\Cinfty(J^1\pi)$ is the {\sl Lagrangian function}.
Using the canonical structures of the bundle $J^1\pi$,
we can construct the {\sl \textbf{Poincar\'e-Cartan $m$-form}} associated with
the Lagrangian density $\Lag$,
denoted by $\Theta_\mathcal{L}\in\Omega^m(J^1\pi)$,
whose local expression is
$$
\Theta_{\Lag}=
\derpar{L}{y^\alpha_i}\,\d y^\alpha\wedge\d^{m-1}x_i-\left(\derpar{L}{y^\alpha_i}y^\alpha_i-L\right)\d^mx \ ,
$$
(where $\dst \d^{m-1}x_i\equiv \inn\left(\derpar{}{x^i}\right)\d^m x$),
and the {\sl \textbf{Poincar\'e-Cartan $(m+1)$-form}}
$\Omega_{\Lag}:=-\d\Theta_\Lag\in\df^{m+1}(J^1\pi)$.
The couple $(J^1\pi,\Omega_\Lag)$ is a {\sl first-order Lagrangian system}
which is said to be {\sl regular} when $\Omega_\Lag$ is $1$-nondegenerate
(that is, a {\sl multisymplectic form\/}) and {\sl singular} elsewhere
(then $\Omega_\Lag$ is {\sl premultisymplectic\/}).
This regularity condition is locally equivalent to
$\displaystyle\det\left(\frac{\partial^2L}
{\partial y^\alpha_i\partial y^\beta_j}(j^1_x\phi)\right)\not= 0$,
for all $j^1_x\phi\in J^1\pi$.

If the theory is second-order
and $\Lag \in \df^{m}(J^2\pi)$ is a second-order Lagrangian density, 
then it is a $\overline{\pi}^2$-semibasic m-form and
$\mathcal{L}=L\,(\overline{\pi}^2)^*\eta$, 
where $L\in\Cinfty(J^2\pi)$ is the Lagrangian function.
As it is well-known, in this case the Lagrangian phase bundle is $J^3\pi$
and natural coordinates on it adapted to the fibration are 
$(x^i,u^\alpha, u_i^\alpha,u_{I}^\alpha,u_{J}^\alpha)$;
$1\leq i\leq m$, $1\leq \alpha \leq n$, and $I$, $J$ are multiindices with $|I|=2$, $|J|=3$.
The {\sl \textbf{Poincar\'e-Cartan $m$-form}} $\Theta_\mathcal{L}\in\Omega^m(J^{3}\pi)$
 for these kinds of theories can be unambiguously constructed 
using again the canonical structures of $J^k\pi$
\cite{art:Aldaya_Azcarraga78_2,proc:Garcia_Munoz83,art:Kouranbaeva_Shkoller00,art:Munoz85,pere}
and it is locally given by
\begin{align*}
\Theta_\Lag &= \left( \derpar{L}{y_i^\alpha} - \sum_{j=1}^{m}\frac{1}{n(ij)} \, \frac{d}{dx^j} \, \derpar{L}{y_{1_i+1_j}^\alpha} \right)(\d y^\alpha \wedge \d^{m-1}x_i - y_i^\alpha\d^mx) \\
&\qquad {} + \frac{1}{n(ij)} \, \derpar{L}{y_{1_i+1_j}^\alpha} \, ( \d y_i^\alpha \wedge \d^{m-1}x_j - y_{1_i+1_j}^\alpha\d^mx ) + L \d^mx \\
&\equiv
L_\alpha^i{\rm d}y^\alpha \wedge{\rm d}^{m-1}x_i+L_\alpha^{ij}{\rm d}y_i^\alpha \wedge{\rm d}^{m-1}x_j +\left(L- L_\alpha^i y_i^\alpha - L_\alpha^{ij}y_{1_i+1_j}^\alpha \right){\rm d}^mx \ ,
\end{align*}
where $L_\alpha^i,L_\alpha^{ij}\in C^{\infty}(J^{3}\pi)$ are
$$
L_\alpha^i=\frac{\partial L}{\partial y_{i}^\alpha} -
 \sum_{j=1}^{m}D_j L^{ij}_{\alpha}\quad ;\quad L_\alpha^{ij}=
\frac{1}{n(ij)}\frac{\partial L}{\partial y_{1_i+1_j}^\alpha} \ .
$$
As above, the {\sl \textbf{Poincar\'e-Cartan $(m+1)$-form}} is
$\Omega_\Lag:=-\d\Theta_\mathcal{L}\in\Omega^{m+1}(J^{3}\pi)$,
so $(J^3\pi,\Omega_\Lag)$ is a {\sl second-order Lagrangian system}
and it is regular or not depending on the $1$-degeneracy of $\Omega_\Lag$.
In this case the regularity condition is locally equivalent to
$\displaystyle\det\left( \derpars{L}{y_I^\beta}{y_J^\alpha} \right)(j^3_x\phi) \neq 0$,
for every $j^3_x\phi\in J^3\pi$, where $|I|=|J|=2$.

The solutions to the Lagrangian variational problem posed by a
first or a second-order Lagrangian $\Lag$
are holonomic sections $j^k\phi\colon M\to J^k\pi$, $k=1,3$, verifying that
\beq
(j^k\phi)^*\inn(X)\Omega_\Lag= 0 \, , \quad \text{for every }X \in \vf(J^k\pi) \, ,
\label{eqsec}
\eeq
or, what is equivalent, they are the integral sections of a class of locally decomposable
non-vanishing, holonomic multivector fields 
$\{ {\bf X}_{\Lag}\}\subset\vf^m(J^k\pi)$, such that
\beq
\inn ({\bf X}_{\Lag})\Omega_{\Lag}=0\ .
 \label{lageqmvf}
\eeq
Holonomic multivector fields are necessarily transverse to the projection 
$\bar\pi^k$ ($k=1,3$) and this condition can be written as
\beq
\inn({\bf X})\omega\neq 0 \ .
 \label{fundeqs}
\eeq
It is usual to fix this condition by taking
a representative in the class $\{{\bf X}\}$ such that
$
\inn({\bf X})\omega=1$; which implies that $f=1$ in \eqref{locholmvf}.

We establish the following notation: let
$\ker^m\Omega_\Lag:=\{{\bf X}\in\vf^m({\cal M})\,\vert\,  \inn({\bf X})\Omega_\Lag=0\}$, 
and let $\ker^m_\omega\Omega_\mathcal{L}$
be the set of $m$-multivector fields satisfying the equation \eqref{lageqmvf} 
and the $\bar\pi^k$-transversality condition \eqref{fundeqs},
but being not necessarily locally decomposable.
Finally, denote by $\ker^m_{\omega(I)}\Omega_\Lag$
the set of integrable $m$-multivector fields satisfying that
they are solutions to  \eqref{lageqmvf} and they are integrable (and holonomic).
Oviously we have
$\ker^m_{\omega(I)}\Omega_\Lag
\subset\ker^m_\omega\Omega_\Lag
 \subset\ker^m\Omega_\Lag$.

\begin{remark}{\rm
\bit
\item
In general, if $(J^k\pi,\Omega_\Lag)$ ($k=1,3$) is a singular Lagrangian system
(i.e., $\Omega_\Lag$ is a premultisymplectic form),
then locally decomposable, non-vanishing, $\bar\pi^k$-transverse multivector fields  which are  solutions 
to the equation \eqref{lageqmvf} could not exist and, in the best of cases, 
they exist only in some submanifold $\jmath_{\cal S}\colon{\cal S}\hookrightarrow J^k\pi$.
This submanifold is necessarily $\bar\pi^k$-transverse,
as a consequence of condition \eqref{fundeqs}.
Furthermore the multivector fields solutions to  \eqref{lageqmvf} 
could not be integrable necessarily (even in the regular case), 
but maybe in some submanifold. 
(An algorithmic procedure in order to find these submanifolds
has been proposed in \cite{LMMMR-2005}).
\item
Notice that, even in the regular case, 
the equation \eqref{lageqmvf} does not determine a unique class of
multivector fields or distributions but a multiplicity of them,
since the solutions depend on arbitrary functions \cite{art:Echeverria_Munoz_Roman98,EMR-99b}.
This means that there is not an unique distribution or a class of multivector fields
solution to \eqref{lageqmvf} on $J^k\pi$ and
then, for every point in $J^k\pi$, there is
a multiplicity of integral submanifolds or integral sections solution to \eqref{eqsec}
(field states) passing through it.
If the Lagrangian system is singular, there is another arbitrariness 
which comes from the degeneracy of the form $\Omega_\Lag$
and is related to the existence of {\sl gauge symmetries},
as we will see in Section \ref{gauge}.
\item
Finally, it is important to point out that integrable and
$\bar\pi^k$-transverse multivector fields ${\bf X}$
solution to the equation\eqref{lageqmvf} may not be necessarily holonomic,
even in the regular case for the second-order case \cite{pere},
although this condition holds in the first-order regular case \cite{art:Echeverria_Munoz_Roman98}.
\eit
}\end{remark}

\section{Symmetries, conservation laws and gauge symmetries}
\label{S3}

\subsection{Symmetries and conserved quantities for Lagrangian field theories}

(See \cite{EMR-99b,art:GPR-2016} for the proofs of all the results in this section. See also \cite{art:deLeon_etal2004,RWZ-2016}).

Let $(J^k\pi,\Omega_\Lag)$ be a Lagrangian system.

\begin{definition}
A \textbf{conserved quantity}
is a form $\xi\in\df^{m-1}(J^k\pi)$ such that
$\Lie({\bf X})\xi:=(-1)^{m+1}\inn({\bf X})\d\xi=0$, for every
 ${\bf X}\in\ker^m_\omega\Omega_\Lag$.
 \end{definition}

The following results characterize conserved quantities:

\begin{theo} 
\ben
\item
 A form $\xi\in\df^{m-1}(J^k\pi)$
 is a conserved quantity if, and only if, $\Lie({\bf Z})\xi=0$, 
 for every ${\bf Z}\in\ker^m\Omega_\Lag$.
\item
 If $\xi\in\df^{m-1}(J^k\pi)$ is a
 conserved quantity 
 and
 ${\bf X}\in\ker^m_{\omega(I)}\Omega_\Lag$,
 then $\xi$ is closed on the integral submanifolds of ${\bf X}$; that is,
if $j_S\colon S\hookrightarrow J^k\pi$
 is an integral submanifold, then $\d j_S^*\xi=0$.
\label{difxi}
\een
 \end{theo}

\begin{remark}{\rm
Given $\xi\in\df^{m-1}(J^k\pi)$ and ${\bf X}\in\vf^m(J^k\pi)$,
for every integral section $\psi\colon M\to J^k\pi$
of ${\bf X}$, there is a unique $X_{\psi^*\xi}\in\vf(M)$ such that
$\inn(X_{\psi^*\xi})\eta=\psi^*\xi$.
This $\psi^*\xi\in\df^{m-1}(M)$ is the so-called {\sl form of flux} associated with the vector field $X_{\psi^*\xi}$
wich is $\psi^*\xi\in\df^{m-1}(M)$ and, if ${\rm div}X_{\psi^*\xi}$ denotes the divergence of $X_{\psi^*\xi}$, we have that
$({\rm div}X_{\psi^*\xi})\,\eta=\d{\psi^*\xi}$.
Then, as a consequence of Proposition \ref{difxi},
$\xi$ is a conserved quantity if,  and only if, ${\rm div}X_{\psi^*\xi}=0$,
and hence, by {\sl Stokes theorem}, in every bounded domain $U\subset M$,
$$
\int_{\partial U}{\psi^*\xi}=\int_U ({\rm div}X_{\psi^*\xi})\,\eta=\int_U\d{\psi^*\xi}=0 \ .
$$
The form $\psi^*\xi$ is called the {\sl \textbf{current}} associated with the conserved quantity $\xi$,
and this result allows to associate 
a {\sl \textbf{conservation law}} in $M$ to every conserved quantity in $J^k\pi$.
}\end{remark}

 \begin{definition}
\ben
\item
A \textbf{symmetry} 
is a diffeomorphism $\Phi\colon J^k\pi
\to J^k\pi$
such that $\Phi_*(\ker^m\Omega_\Lag)\subset\ker^m\Omega_\Lag$.

If $\Phi=j^k\varphi$ \  for a diffeormorphism $\varphi\colon E\to E$, 
the symmetry is called \textbf{natural}.
\item
 An \textbf{infinitesimal symmetry}
is a vector field $Y\in\vf ({\cal M})$ whose local flows are 
local symmetries or, what is equivalent, such that \
$[Y,\ker^m\Omega_\Lag]\subset\ker^m\Omega_\Lag$.

If $Y=j^kZ$ for some 
$Z\in\vf (M)$, then the infinitesimal symmetry is called \textbf{natural}.
\een
 \end{definition}

For infinitesimal symmetries we also have the following characterization:

 \begin{theo}
$Y\in\vf(J^k\pi)$ is an infinitesimal symmetry if, and only if,
$[Y,\ker^m\Omega_\Lag]\subset\ker^m\Omega_\Lag$.
 \label{previo0}
 \end{theo}

Observe that, if $Y_1,Y_2\in\vf ({\cal M})$ are infinitesimal symmetries,
 then so is $[Y_1,Y_2]$.
Furthermore, if $Y\in\vf ({\cal M})$ is an infinitesimal symmetry
then, for every $Z\in\ker\Omega_\Lag$,
$Y+Z$ is also an infinitesimal symmetry.

The Lagrangian field equations are EDP's and symmetries transform solutions into solutions.
In fact:

\begin{theo}
 Let $\Phi\in{\rm Diff}(J^k\pi)$ be a symmetry.
Then:
 \ben
 \item
 For every integrable multivector field ${\bf X}\in\ker^m\Omega_\Lag$, 
 the map $\Phi$ transforms integral submanifolds
 of ${\bf X}$ into integral submanifolds of $\Phi_*{\bf X}$.
 \item
 In the particular case that \,$\Phi\in{\rm Diff}(J^k\pi)$ restricts to a diffeormorphism 
$\varphi\colon M\to M$ (which means that $\varphi\circ\bar\pi^k=\bar\pi^k\circ\Phi$);
then, for every ${\bf X}\in\ker^m_{\omega(I)}\Omega_\Lag$,
the map $\Phi$ transforms integral submanifolds of ${\bf X}$
into integral submanifolds of \,$\Phi_*{\bf X}$, and hence
$\Phi_*{\bf X}\in\ker^m_{\omega(I)}\Omega_\Lag$.
 \een
 \label{gsymsol}
 \end{theo}

As a straightforward consequence of this, we obtain that:

 \begin{theo}
 Let $Y\in\vf (J^k\pi)$ be an infinitesimal symmetry
and $F_t$ the local flow of $Y$. Then:
 \ben
 \item
 For every integrable multivector field ${\bf X}\in\ker^m\Omega_\Lag$, the map $F_t$ transforms integral submanifolds
 of ${\bf X}$ into integral submanifolds of $F_{t*}{\bf X}$.
 \item
 In the particular case that $Y\in\vf (J^k\pi)$ is $\bar\pi^k$-projectable
(which means that there exists $Z\in\vf (M)$ such that
the local flows of $Z$ and $Y$ are $\bar\pi^k$-related);
then, for every ${\bf X}\in\ker^m_{\omega(I)}\Omega_\Lag$,
$F_t$ transforms integral submanifolds of ${\bf X}$
into integral submanifolds of $F_{t*}{\bf X}$, and hence
$F_{t*}{\bf X}\in\ker^m_{\omega(I)}\Omega_\Lag$.
 \een
 \label{gsymsol2}
 \end{theo}

If $\Phi\in{\rm Diff}(J^k\pi)$ is a symmetry and $\xi\in\df^{m-1}(J^k\pi)$
is a conserved quantity,  then $\Phi^*\xi$ is also a conserved quantity.
In the same way, if $Y\in\vf (J^k\pi)$ is an infinitesimal symmetry and 
$\xi\in\df^{m-1}(J^k\pi)$ is a conserved quantity,
then $\Lie(Y)\xi$ is also a conserved quantity.

The most relevant kinds of symmetries a Lagrangian system
$(J^k\pi,\Omega_\Lag)$ are the following:

 \begin{definition}
\ben
\item
A \textbf{Cartan} or \textbf{Noether symmetry} 
is a diffeomorphism $\Phi\colon J^k\pi\to J^k\pi$
such that, $\Phi^*\Omega_\Lag=\Omega_\Lag$.
If, in addition, $\Phi^*\Theta_\Lag=\Theta_\Lag$, then $\Phi$ is said to be an
\textbf{exact Cartan} or \textbf{Noether symmetry}.

If $\Phi=j^k\varphi$ \ 
for a diffeormorphism $\varphi\colon E\to E$, 
the Cartan symmetry is called \textbf{natural}.
\item
An \textbf{infinitesimal Cartan} or \textbf{Noether symmetry} 
is a vector field $Y\in\vf (J^k\pi)$ satisfying that $\Lie(Y)\Omega_\Lag=0$.
If, in addition, $\Lie(Y)\Theta_\Lag=0$, then $Y$ is said to be an
\textbf{infinitesimal exact Cartan} or \textbf{Noether symmetry}.

If $Y=j^kZ$ for some 
$Z\in\vf (E)$, then the infinitesimal Cartan symmetry is called \textbf{natural}.
\een
 \end{definition}

Obviously, if $Y_1,Y_2\in\vf (J^k\pi)$
are infinitesimal Cartan or Noether symmetries, then so is $[Y_1,Y_2]$.

Now, if $\psi\colon M\to J^k\pi$ ($k=1,3$)
is a solution to the equation \eqref{eqsec} and 
\,$\Phi\in{\rm Diff}(J^k\pi)$ is a Cartan or Noether symmetry,
then, for every $X\in\vf(J^k\pi)$, we have (see Theorem \ref{gsymsol})
\beq
(\Phi\circ\psi)^*\inn(X)\Omega_\Lag=
\psi^*\Phi^*\inn(X)\Omega_\Lag=
\psi^*\inn(\Phi_*^{-1}X)(\Phi^*\Omega_\Lag)=
\psi^*\inn(X')\Omega_\Lag=0 \ ,
\label{holcons0}
\eeq
since $X'=\Phi_*^{-1}X\in\vf(J^k\pi)$, $\Phi^*\Omega_\Lag=\Omega_\Lag$, 
and $\Phi^*\Omega_\Lag=0$.
Therefore $\Phi\circ\psi$ is also a solution to \eqref{eqsec};
thus $\Phi$ transforms solutions into solutions and then it is a symmetry.

Furthermore, if $\psi=j^k\phi\colon M\to J^k\pi$ ($k=1,3$)
is a holonomic solution to the equation \eqref{eqsec} and
$\Phi=j^k\varphi\in{\rm Diff}(J^k\pi)$ is a natural Cartan or Noether symmetry,
then \eqref{holcons0} reads
\bea
(j^k(\varphi\circ\phi))^*\inn(X)\Omega_\Lag&=&
(j^k\phi)^*(j^k\varphi)^*\inn(X)\Omega_\Lag
\nonumber \\ &=&
(j^k\phi)^*\inn((j^k\varphi)_*^{-1}X)((j^k\varphi)^*\Omega_\Lag)=
(j^k\phi)^*\inn(X')\Omega_\Lag=0 \ ,
\label{holcons2}
\eea
and therefore $j^k(\varphi\circ\phi)$ is also a 
holonomic solution to \eqref{eqsec}.
Thus $\Phi=j^k\varphi$ transforms holonomic solutions into holonomic solutions.

In addition, if $Y\in\vf({\cal M})$ is an infinitesimal (natural) Cartan or Noether symmetry,
by definition, its local flows are local (natural) Cartan or Noether symmetries.

In this way we have proved that:

\begin{prop}
Every Cartan or Noether symmetry is a symmetry and, as a consequence,
every infinitesimal Cartan or Noether symmetry is an infinitesimal symmetry.

Furthermore, every natural (infinitesimal) Cartan symmetry transforms holonomic solutions to the field equations into holonomic solutions.
\label{symsect}
\end{prop}

The condition $\Lie(Y)\Omega_\Lag=0$
 is equivalent to demanding that $\inn(Y)\Omega$
 is a closed $m$-form in $J^k\pi$.
 Thus, an infinitesimal Cartan or Noether symmetry is a {\sl locally Hamiltonian vector field}
 for the multisymplectic form $\Omega_\Lag$,
 and $\xi_Y$ is the corresponding {\sl local Hamiltonian form},
(in an open neighbourhood of every point in $J^k\pi$).
Therefore, {\sl Noether's theorem} is stated as follows:

 \begin{theo}
 {\rm (Noether):}
Let $Y\in\vf (J^k\pi)$ be an infinitesimal Cartan or Noether symmetry
with $\inn(Y)\Omega_\Lag=\d\xi_Y$
in an open set $U\subset J^k\pi$. Then,
 for every ${\bf X}\in\ker^m_\omega\Omega_\Lag$
(and hence for every ${\bf X}\in\ker^m_{\omega(I)}\Omega_\Lag$), we have
 $$
 \Lie({\bf X})\xi_Y=0 \ ;
 $$
 that is, any Hamiltonian $(m-1)$-form
 $\xi_Y$ associated with $Y$ is a conserved quantity.
Then, in this context, for every integral submanifold $\psi$ of ${\bf X}$,
 the form $\psi^*\xi_Y$ is usually called a {\sl \textbf{Noether current}}.
 \label{Nth}
 \end{theo}

Observe that the form $\Lie(Y)\Theta_\Lag$ is closed since
 $$
 \Lie(Y)\Theta_\Lag=\d\inn(Y)\Theta_\Lag+\inn(Y)\d\Theta_\Lag=
\d\inn(Y)\Theta_\Lag-\inn(Y)\Omega_\Lag=
\d (\inn(Y)\Theta_\Lag-\xi_Y)\equiv\d\zeta_Y
 \quad \mbox{\rm (in $U$)} \ .
 $$
In particular, if $Y$ is an exact infinitesimal Cartan or Noether symmetry, we can take
 $\xi_Y=\inn(Y)\Theta_\Lag$.

It is well known that canonical liftings of diffeomorphisms and vector
fields preserve the canonical structures of $J^k\pi$. Nevertheless,
the (pre)multisymplectic form $\Omega_\Lag$ is not
canonical, since it depends on the choice of the Lagrangian density
$\Lag$, and then it is not invariant by these canonical liftings. Thus,
given a diffeomorphism $\Phi\colon J^k\pi\to J^k\pi$ or a vector
field $Y\in\vf (J^k\pi)$, a sufficient condition to assure this invariance
would be to demand that they leave the canonical
structures of the jet bundle $J^k\pi$ (for instance,
$\Phi$ and $Y$ being the canonical lifting of a diffeomorphism and a
vector field in $E$), and that the Lagrangian density $\Lag$ be also
invariant. In this way, $\Omega_\Lag$ and hence the
Euler-Lagrange equations are invariant by $\Phi$ or $Y$.
This leads to define the following kind of symmetries:

\begin{definition}
\label{gaugelag}
\ben
\item
A \textbf{Lagrangian symmetry} of the Lagrangian system is a diffeomorphism
$\Phi\colon J^k\pi\to J^k\pi$ such that:
\ben
\item
$\Phi$ leaves the canonical geometric structures of $J^k\pi$ invariant.
\item
$\Phi^*\Lag=\Lag$ ($\Phi$ leaves $\Lag$ invariant).
\een
If $\Phi=j^k\varphi$, for some
diffeomorphism $\varphi\colon E\to E$, then condition (a) holds and
the Lagrangian symmetry is called \textbf{natural}.
\item
An \textbf{infinitesimal Lagrangian symmetry} is a
vector field $Y\in\vf (J^k\pi)$ such that:
\ben
\item
The canonical geometric structures of $J^k\pi$ are invariant
under the action of $Y$.
\item
 $\Lie(Y)\Lag =0$ ($Y$ leaves $\Lag$ invariant).
\een
If $\Phi=j^k\varphi$, for some
diffeomorphism $\varphi\colon E\to E$, then condition (a) holds and
the infinitesimal Lagrangian symmetry is called \textbf{natural}.
\label{sym}
\een
\end{definition}

As a direct consequence of these definitions we have:

\begin{prop}\label{prop:jet}
\ben
\item
If $\Phi\in{\rm Diff}(J^k\pi)$ is a Lagrangian symmetry, then
$\Phi ^*\Theta_{\Lag} =\Theta_{\Lag}$, and hence
it is an exact Cartan symmetry.
\item
If \ $Y\in\vf (J^k\pi)$ is an infinitesimal Lagrangian symmetry, then
$\Lie (Y)\Theta_{\Lag} =0$, and hence it is
an infinitesimal exact Cartan symmetry.
\een
\label{invasim}
\end{prop}

To demand the invariance of $\Lag$ is really a strong condition, since there are
Lagrangian densities or, what is equivalent, Lagrangian functions that, 
being different and even of different order,  give rise to the 
same Euler-Lagrange equations.
These are the so-called {\sl gauge equivalent Lagrangians}.

\subsection{Gauge symmetries, gauge vector fields and gauge equivalence}
\label{gauge}

The term {\sl gauge} is used in Physics to refer to different situations
in relation to certain kinds of symmetries which do not change physically the system,
and this characteristic is known as {\sl gauge freedom}.
For instance, sometimes it is used to refer to the invariance of a Lagrangian system 
when is described by different Lagrangian functions which lead to the same
Euler-Lagrange equations (really these are the Lagrangian symmetries
introduced in Definition \ref{gaugelag}).
Nevertheless, the standard use of the concept ``gauge'' is for describing
symmetries related to the non-regularity of the Lagrangian
and lead to the existence of states that are physically equivalent.
From a geometric point of view, these kinds of symmetries
are closely related with the degeneracy of the
Poincar\'e--Cartan forms associated with the Lagrangians.
In this section we introduce and analyze the geometric concept of these {\sl gauge symmetries\/}
for Lagrangian field theories.
This discussion is inspired in the geometric treatment made mainly in
\cite{BK-86,GN-79} about {\sl gauge vector fields} and 
{\sl gauge equivalent states}
for non-regular dynamical systems.

Consider a singular Lagrangian system $(J^k\pi,\Omega_\Lag)$ and
assume that equations \eqref{lageqmvf} have solutions on a 
$\bar\pi^k$-transverse submanifold 
${\rm j}_{\cal S}\colon{\cal S}\hookrightarrow J^k\pi$ 
(it could be ${\cal S}=J^k\pi$).
As we have said, the existence of gauge symmetries or {\sl gauge freedom}
which we are interested here
is closely related with the fact that the Lagrangian theory is
non-regular; that is, the Poincar\'e-Cartan form $\Omega_\Lag$
is $1$-degenerated and then it is a premultisymplectic form.
As a consequence,
besides to the non unicity of solutions which is characteristic of classical field theories,
there is another family of additional solutions which is associated with this degeneracy.

The local generators of gauge symmetries are called {\sl gauge vector fields}.
In order to do a more accurate geometric definition,
let $\underline{\vf({\cal S})}\subset\vf(J^k\pi)$ be the set of vector fields in $J^k\pi$ 
which are tangent to the submanifold ${\cal S}$,
and let
$$
\underline{\ker\,\Omega_\Lag}:=\{ Z\in\underline{\vf({\cal S}})\,\vert\, {\rm j}_{\cal S}^*\inn(Z)\Omega_\Lag=0\} \ ,
$$
or, what is equivalent, if $Z^{\cal S}\in\vf({\cal S})$
is such that ${\rm j}_{{\cal S}*}Z^{{\cal S}}=Z\vert_{\cal S}$,
for every $Z\in\underline{\vf({\cal S})}$; then
$$
0={\rm j}_{\cal S}^*\inn(Z)\Omega_\Lag=\inn(Z^{{\cal S}})({\rm j}_{\cal S}^*\Omega_\Lag) \ ,
$$
and hence $Z^{\cal S}\in\ker\,\Omega_\Lag$.
Denote by $\vf^{V(\bar\pi^k)}(J^k\pi)$ the set of 
$\bar\pi^k$-vertical vector fields in $J^k\pi$ and let 
$\ker^{V(\bar\pi^k)}\Omega_\Lag=\ker\Omega_\Lag\cap\vf^{V(\bar\pi^k)}(J^k\pi)$.
Finally consider the set
$$
{\cal G}=\ker^{V(\bar\pi^k)}\Omega_\Lag\cap\underline{\vf({\cal S})}
$$
(the $\bar\pi^k$-vertical vector fields of $\ker\Omega_\Lag$ 
which are tangent to ${\cal S}$).
Therefore, gauge vector fields must have the following properties:
\bit
\item
As physical states are sections of the projection $\bar\pi^k$ with image on ${\cal S}\subseteq J^k\pi$,
gauge vector fields must be $\overline{\pi}^k$-vertical.
In this way, we assure that the base manifold $M$ 
does not contain gauge equivalent points and then all the 
gauge degrees of freedom are in the fibres of $J^k\pi$ and,
therefore, after removing the gauge redundancy,
the base manifold $M$ remains unchanged.
\item
As the flux of gauge vector fields connect equivalent physical states,
they must be tangent to ${\cal S}$; that is, elements of $\underline{\vf({\cal S})}$.
\item
As the existence of gauge symmetries
is a consequence of the non-regularity of the Lagrangian $\Lag$ (and conversely);
gauge vector fields are necessarily
related with the premultisymplectic character of the
Poincar\'e-Cartan form $\Omega_\Lag$. 
Hence, they should be elements of the set
$\underline{\ker\,\Omega_\Lag}$.
The flux of these vector fields transforms solutions to the
Lagrangian field equations into solutions but,
in principle, without preserving the holonomy necessarily.
\item
It is also usual to demand that physical symmetries are
{\sl natural}. This means that they are canonical liftings to the 
bundle of phase states of symmetries in the configuration space $E$; 
that is, canonical lifts to $J^k\pi$ of vector fields in $E$.
Then, by Proposition \ref{symsect}, this condition assures that gauge symmetries transform 
holonomic solutions to the field equations into holonomic solutions.
\eit

These ideas lead to state the following definitions:

\begin{definition}
The elements  \,$Z\in{\cal G}$ 
are called \textbf{gauge vector fields} or \textbf{(infinitesimal) gauge symmetries}
of the singular Lagrangian system $(J^k\pi,\Omega_\Lag)$.
If $Z$ is the canonical lift of a vector field in $E$,
then it is called a \textbf{natural gauge vector field}.
\label{gaugevectors}
\end{definition}

As ${\cal G}\subset\ker\Omega_\Lag$, for every $Z\in{\cal G}$ we have that
$\inn(Z)\Omega_\Lag=0$, then $\Lie(Z)\Omega_\Lag=0$.
Therefore every gauge vector field is an infinitesimal Cartan symmetry
and then a symmetry (Proposition \ref{symsect}),
and hence, if it is also a holonomic vector field,
it  transforms holonomic solutions to the field equations into holonomic solutions (see \eqref{holcons2}). Incidentally, any  closed $m-1$-forms can be thought as an associated local Hamiltonian form to any gauge vector field. 
Furthermore, gauge vector fields are trivially $\bar\pi^k$-projectable 
(to the null vector field on $M$); therefore the item 2 in Theorem \ref{gsymsol2} holds.

Observe that, for every $Z_1,Z_2\in{\cal G}$ we have that $[Z_1,Z_2]\in{\cal G}$,
and hence ${\cal G}$ generates an involutive distribution. Then, we construct the quotient set $\widetilde{\cal S}={\cal S}/{\cal G}$.

\begin{definition}
\begin{itemize}
    \item  Points of $\cal S$ which are in the same class in $\widetilde{\cal S}$
are said to be \textbf{gauge equivalent points}.
\item Two sections $\psi_1,\psi_2:M\rightarrow \cal S$ are \textbf{gauge equivalent} if $\psi_1(x)$ is gauge equivalent to $\psi_2(x)$ for any $x\in M$.
\item If $\psi_1,\psi_2$ are gauge equivalent and solutions to the field equations \eqref{eqsec}, they are also called
\textbf{gauge equivalent field states}.
\end{itemize}
\end{definition}

Therefore, the following statement is assumed:

\noindent {\bf Statement}: {\it
\textbf{(Gauge principle)}.
Gauge equivalent field states are physically equivalent or, what means the same thing,
they represent the same physical state of the field.}

As a consequence of the gauge principle,
we can choose any representative in each gauge equivalence class
of sections to represent a physical state.
This is known as a {\sl gauge fixing} and
the freedom in choosing the representative  is referred as the 
{\sl gauge freedom} of the theory. Notice that a holonomic section could be gauge equivalent to a non-holonomic one. It may happen that that there is only one holonomic representative in a class.

When, as a consequence of the degeneracy of the Lagrangian,
a Lagrangian system has gauge symmetries, a relevant problem consists in removing the unphysical redundant information
introduced by the existence of gauge equivalent states.
This can be done by implementing 
the well-known procedure of  {\sl reduction by symmetries}
which rules as follows:
since ${\cal G}$ generates an involutive distribution, we construct the quotient set $\widetilde{\cal S}={\cal S}/{\cal G}$,
which is assumed to be a differentiable manifold which is made of the true physical degrees of freedom.
In addition, $\widetilde\pi_{\cal S}$ is a fiber bundle over $M$,
and the `real' physical states are the sections of the projection 
$\tilde\pi_{\cal S}\colon\widetilde{\cal S}\to M$.
 We have the diagram
$$
\xymatrix{
{\mathcal{S}} \ar[drr]^-{\pi_{\cal S}} \ar[rr]^-{{\jmath_{\cal S}}} \ar[d]^-{\tilde{\tau}_{\cal S}} & \ & J^k\pi
 \ar[d]^-{\bar\pi^k}  \\
\widetilde{\mathcal{S}} \ar[rr]^-{\tilde\pi_{\cal S}} & \ & M
} \ .
$$
This is what is known as the {\sl \textbf{gauge reduction procedure}} for removing the (unphysical)
gauge degrees of freedom of the theory.
An alternative way to remove the gauge freedom consists in taking 
a (local) section of the projection 
$\tilde\tau_{\cal S}\colon{\cal S}\to\widetilde{\cal S}$,
and this is what is called a {\sl \textbf{gauge fixing}}.

\begin{remark}{\rm
To ensure that the base manifold $M$ does not contain gauge equivalent points 
(that is, that all the gauge degrees of freedom are in the fibers of $J^k\pi$),
we have demanded that gauge vector fields are only the 
$\bar\pi^k$-vertical vector fields of $\ker\Omega_\Lag$,
and not all the elements of this set.
In this way, after doing this reduction procedure
or a gauge fixing in order to remove the gauge redundancy,
the base manifold $M$ remains unchanged.}
\end{remark}

As it has been commented, the existence of gauge symmetries and of
gauge freedom is a consequence
of the non-regularity of the Lagrangian $\Lag$ (and conversely);
and then it is related with the premultisymplectic character of the form $\Omega_\Lag$.
Then, it is reasonable to think that the gauge reduction procedure, which removes 
the (unphysical) gauge degrees of freedom, 
must remove also the degeneracy of the form.
In order to analyze this question, we have to consider the form
$\Omega_\Lag^{\cal S}=\jmath_{\cal S}^*\Omega_\Lag$. Then we denote 
$$
\underline{\ker\Omega_\Lag^{\cal S}}:=\{ Z\in\underline{\vf({\cal S})}\ \vert\ 
\exists Z\in\ker\Omega_\Lag\ \ \vert \ \ \jmath_{{\cal S}*}Z=Z\vert_{\cal S}\} \ .
$$
It is obvious that 
${\cal G}_1\subseteq\underline{\ker\Omega_\Lag^{\cal S}}$
since, for every $Z\in{\cal G}_1\subset\ker\,\Omega_\Lag$,
$$
\inn(Z)\Omega_\Lag=0 \ \Longrightarrow\ 
0=\jmath_{\cal S}^*\inn(Z)\Omega_\Lag=\inn(Z_{\cal S})\Omega_\Lag^{\cal S} 
 \ \Longrightarrow\  Z_S\in \ker\,\Omega_\Lag^{\cal S}
\ \Longleftrightarrow\  Z\in\underline{\ker\,\Omega_\Lag^{\cal S}}\ .
$$
(Observe that, if $\Omega_\Lag^{\cal S}$ is nondegenerate (multisymplectic),
then $\underline{\ker\,\Omega_\Lag^{\cal S}}=\{ 0\}$,
which implies that $\ker\Omega_\Lag\cap\underline{\vf({\cal S})}=\{ 0\}$
and then ${\cal G}_1=\{ 0\}$).
As $\Omega_\Lag^{\cal S}$ is a closed form, we have also that
$\Lie(Z)\Omega_\Lag=0$, for every $Z\in{\cal G}_1$;
then it is $\tilde\tau_{\cal S}$-projectable to a form 
$\widetilde\Omega_\Lag^{\cal S}\in\df^m(\widetilde{\cal S})$.
Therefore, to say that the gauge reduction procedure 
consisting in making the quotient of 
${\cal S}$ by ${\cal G}_1$  removes the degeneracy
is equivalent to say that $\widetilde\Omega_\Lag^{\cal S}$ is a multisymplectic form,
and this happens if, and only if, ${\cal G}_1=\underline{\ker\Omega_\Lag^{\cal S}}$.
In general, this last condition does not hold and, in this case, 
if we want that the gauge reduction removes the degeneracy,
we need to enlarge the set of admisible gauge vector fields.
So, by similarity with the case of presmplectic mechanics
\cite{BK-86,GN-79}, we can define:

\begin{definition}
${\cal G}=\underline{\ker\Omega_\Lag^{\cal S}}\cap\vf^{V(\bar\pi^k)}(J^k\pi)$ 
is the \textbf{complete set of gauge vector fields} 
for the singular Lagrangian system $(J^k\pi,\Omega_\Lag)$.
Then, the elements  
\,$Z\in{\cal G}_1$  are called \textbf{primary gauge vector fields} and those
$Z\in{\cal G}-{\cal G}_1$ are called  \textbf{secondary gauge vector fields}.
\end{definition}

As above we have forced gauge vector fields to be $\bar\pi^k$-vertical.
This means that, unless ${\cal G}=\underline{\ker\Omega_\Lag^{\cal S}}$,
the gauge reduction procedure do not remove entirely the degeneracy of the form
$\Omega_\Lag$.

\begin{remark}{\rm
In the particular situation where $(J^k\pi,\Omega_\Lag)$ 
is a singular Lagrangian system such that 
the equations \eqref{lageqmvf} have solutions on $J^k\pi$,
then the gauge vector fields are all the elements of $\ker^{V(\bar\pi^k)}\Omega_\Lag$,
since $[Z_1,Z_2]\in\ker\,\Omega_\Lag^{V(\bar\pi^k)}$, 
for every $Z_1,Z_2\in\ker\,\Omega_\Lag^{V(\bar\pi^k)}$.
Then, when $\ker^{V(\bar\pi^k)}\Omega_\Lag=\ker\,\Omega_\Lag$,
the reduction procedure removes both the unphysical degrees of freedom 
and the degeneracy of the premultisymplectic structure.
If $\ker^{V(\bar\pi^k)}\Omega_\Lag\subset\ker\,\Omega_\Lag$,
the vector fields of $\ker\,\Omega_\Lag$ which are not $\bar\pi^k$-vertical
would not be gauge vector fields, but just (infinitesimal) Noether symmetries.}
\end{remark}


\section{Electromagnetic field}
\label{EM}

Given the $G$-principle bundle $P\rightarrow M$, with $M$ a $4$-dimensional manifold and group $G=U(1)$, consider the connection bundle $\pi\colon C\rightarrow M$, 
with local adapted coordinates $(x^\alpha,A_\alpha)$, where $\alpha=0,\dots,3$.
The induced coordinates in the  associated first-order jet bundle $J^1\pi$ are 
$(x^\alpha,A_\alpha,A_{\alpha,\beta})$, where $A_\alpha$ denote the components
of the {\sl potential of the electromagnetic field} $A$.
The Maxwell Lagrangian at vacuum is (taking the magnetic constant $\mu_0=1$)
$$
L=\frac14 F^{\alpha\beta}F_{\alpha\beta}=\frac14(A^{\beta,\alpha}-A^{\alpha,\beta})(A_{\beta,\alpha}-A_{\alpha,\beta})=\frac12(\eta^{\alpha\nu}\eta^{\beta\mu}-\eta^{\alpha\mu}\eta^{\beta\nu})A_{\alpha,\beta}A_{\mu,\nu} \ ;
$$
$\eta^{\alpha\beta}$ are the components of the inverse of the Minkowski metric. The associated Poincar\'e-Cartan form associated with the Lagrangian density ${\cal L}=L\, \d^4x$ is
$$
\Omega_\Lag=\eta^{\alpha\mu}\eta^{\beta\nu}F_{\alpha\beta}\d A_{\mu,\nu}\wedge \d^4x-(\eta^{\alpha\nu}\eta^{\beta\mu}-\eta^{\alpha\mu}\eta^{\beta\nu})\d A_{\mu,\nu}\wedge\d A_{\alpha}\wedge\d^3x_\beta \ .
$$
A general locally decomposable multivector field has the local expression
$$
{\bf X}=
\bigwedge^3_{\gamma=0}\left(\derpar{}{x^\gamma}+
f_{\alpha,\gamma}\derpar{}{A_\alpha}+
G_{\alpha\beta,\gamma}\derpar{}{A_{\alpha,\beta}}\right) \ ,
$$
and the field equation \eqref{lageqmvf} reads as
$$
(\eta^{\alpha\nu}\eta^{\beta\mu}-\eta^{\alpha\mu}\eta^{\beta\nu})(A_{\alpha,\beta}-f_{\alpha\beta})=0 \quad , \quad
(\eta^{\alpha\nu}\eta^{\beta\mu}-\eta^{\alpha\mu}\eta^{\beta\nu})G_{\mu\nu,\beta}=G^{\mu\beta}_\mu-G^{\beta\mu}_\mu=0 \ . 
$$
The first group of equations implies $f_{\alpha\beta}=A_{\alpha,\beta}+T_{\alpha\beta}$, 
where $T_{\alpha\beta}-T_{\beta\alpha}=0$. When we impose holonomy we deduce $T_{\alpha\beta}=0$. 
The second group of equations, for the integral sections of ${\bf X}$, 
leads to Maxwell's equations
$$
\frac{\partial^2A^\beta}{\partial x^\mu\partial x_\mu}-
\frac{\partial^2A^\mu}{\partial x^\mu\partial x_\beta}=0 \ ;
$$
and there are no constraints, so ${\cal S}=J^1\pi$.
The gauge vector fields are
$$
{\cal G}=
\left\{S_{\alpha\beta}\derpar{}{A_{\alpha,\beta}}\ | \ S_{\alpha\beta}\in C^\infty(J^1\pi) \text{ , such that } S_{\alpha\beta}-S_{\beta\alpha}=0\right\} \ .
$$

Two sections are gauge equivalent if $(x^\mu,A_\alpha(x),A_{\alpha,\beta}(x))=(x^\mu,A'_\alpha(x),A'_{\alpha,\beta}(x)+s_{\alpha\beta}(x))$, for some set of functions $s_{\alpha\beta}(x)$ which are symmetric by the interchange of $\alpha$ and $\beta$. 
In particular, $A_\alpha(x)=A'_\alpha(x)$; therefore, if both sections are holonomic, $s_{\alpha\beta}(x)=0$. In other words, there is only one holonomic section in every gauge equivalent class.

This result may confront the well know physical result: 
for every section $B_\alpha(x^\mu)$
solution to the field equation \eqref{eqsec}, 
we can find another solution by the transformation 
$\dst B'_\alpha(x^\mu)=
B_\alpha(x^\mu)+\derpar{f(x^\mu)}{x^\alpha}$,
for any function $f$. This induces the transformation 
\begin{align*}
   \Psi:J^1\pi&\rightarrow J^1\pi 
   \\
   (x^\mu,A_\alpha,A_{\alpha,\beta})&\mapsto \left(x^\mu,A_\alpha+\derpar{f}{x^\alpha},A_{\alpha,\beta}+\frac{\partial^2f}{\partial x^\alpha\partial x^\beta}\right).
\end{align*}
 This transformation is actually a Lagrangian symmetry.


\section{Gravitational field (General Relativity)}
\label{S4}

The multisymplectic approach to the Einstein--Hilbert and the
Einstein--Palatini models of General relativity
has been done, for instance, in \cite{first,art:Capriotti,art:Capriotti2,GR1,GR2,GIMMSY,vey1}
(see also the references therein).

\subsection{The Hilbert-Einstein action}

Fist we consider the Hilbert Lagrangian for the Einstein equations of gravity
without sources (no matter-energy is present).

The configuration bundle for the system is $\pi\colon E\rightarrow M$, 
where $M$ is a connected 4-dimensional manifold representing space-time and
$E$ is the manifold of Lorentzian metrics on $M$;
that is, for every $x\in M$, the fiber $\pi^{-1}(x)$ 
is the set of metrics acting on $\Tan_xM$, 
with signature $(1,3)$ (i.e.; $(-+++)$).
Local coordinates in $E$ are denoted $(x^\mu,g_{\alpha\beta})$,
 with $0\leq\alpha\leq\beta\leq3$. As $g$ is symmetric,
$g_{\alpha\beta}=g_{\beta\alpha}$, actually there are 10 independent variables and,
hence, the dimension of the fibers is $10$ and $\dim E=14$.
(The fact that $g$ is a Lorentz metric is not explicitly shown and
is included requiring that the Lagrangian is invariant under Lorentz transformations). 
The induced coordinates in $J^3\pi$ are 
$(x^\mu,\,g_{\alpha\beta},\,g_{\alpha\beta,\mu},\,g_{\alpha\beta,\mu\nu},
\,g_{\alpha\beta,\mu\nu\rho})$.

Using these coordinates, the local expression of the Hilbert-Einstein Lagrangian is
$$
L_{EH}=\sqrt{|{\rm det}(g)|}\,R=
\sqrt{|{\rm det}(g)|}\,g^{\alpha\beta}R_{\alpha\beta}\equiv
\varrho g^{\alpha\beta}R_{\alpha\beta}=
\varrho R	\ ,
$$
where $\varrho\equiv\sqrt{|det(g_{\alpha\beta})|}$,
$R=g^{\alpha\beta}R_{\alpha\beta}$ is the {\sl scalar curvature},
$R_{\alpha\beta}=D_\gamma\Gamma^{\gamma}_{\alpha\beta}-D_\alpha\Gamma^{\gamma}_{\gamma\beta}+
\Gamma^{\gamma}_{\alpha\beta}\Gamma^{\delta}_{\delta\gamma}-
\Gamma^{\gamma}_{\delta\beta}\Gamma^{\delta}_{\alpha\gamma}$
are the components of the {\sl Ricci tensor},
$\displaystyle\Gamma^{\rho}_{\mu\nu}=
\frac{1}{2}\,g^{\rho\lambda}\left(\frac{\partial g_{\nu\lambda}}{\partial x^\mu}+ 
\frac{\partial g_{\lambda\mu}}{\partial x^\nu}- \frac{\partial g_{\mu\nu}}{\partial x^\lambda}\right)=
\frac{1}{2}\,g^{\rho\lambda}(g_{\nu\lambda,\mu}+ 
g_{\lambda\mu,\nu}-g_{\mu\nu,\lambda})$ 
are the {\sl Christoffel symbols of
the Levi-Civita connection} of $g$, and
$g^{\alpha\beta}$ denotes the inverse matrix of $g$, 
namely: $g^{\alpha\beta}g_{\beta\gamma}=\delta^\alpha_\gamma$.
It is useful to consider the following decomposition \cite{first,rosado}:
$$
L_{EH}=\sum_{\alpha\leq\beta}L^{\alpha\beta,\mu\nu}g_{\alpha\beta,\mu\nu}+L_0 \ ,
$$
where
\beann
L^{\alpha\beta,\mu\nu}&=&
\frac{1}{n(\mu\nu)}\frac{\partial L}{\partial g_{\alpha\beta,\mu\nu}}=
\frac{n(\alpha\beta)}{2}\varrho(g^{\alpha\mu}g^{\beta\nu}+g^{\alpha\nu}g^{\beta\mu}-2g^{\alpha\beta}g^{\mu\nu})\ ,
 \\ 
L_0&=&\varrho g^{\alpha\beta}\{g^{\gamma\delta}(g_{\delta\mu,\beta}\Gamma^{\mu}_{\alpha\gamma}-g_{\delta\mu,\gamma}\Gamma^{\mu}_{\alpha\beta}) +\Gamma^{\delta}_{\alpha\beta}\Gamma^{\gamma}_{\gamma\delta}-\Gamma^{\delta}_{\alpha\gamma}\Gamma^{\gamma}_{\beta\delta}\}\ .
\eeann
The key point on this decomposition is that $L^{\alpha\beta,\mu\nu}$ and $L_0$
project onto functions of $C^\infty(E)$ and 
$C^\infty(J^1\pi)$, respectively.

The Poincar\'e-Cartan $3$-form $\Theta_{\mathcal{L}_{EH}}$ 
associated with the Hilbert-Einstein Lagrangian density 
$\mathcal{L}_{EH}=
L_{EH}\,(\overline{\pi}^3)^*\eta=L_{EH}\,\d^4x$ is
$$
\Theta_{\mathcal{L}_{EH}}=
-H\,\d^4x
+\sum_{\alpha\leq\beta}L^{\alpha\beta,\mu}\d g_{\alpha\beta}\wedge \d^{m-1}x_\mu+\sum_{\alpha\leq\beta}L^{\alpha\beta,\mu\nu}\d g_{\alpha\beta,\mu}\wedge \d^{m-1}x_{\nu}  \ ;
$$
where
\beann
L^{\alpha\beta,\mu}&=&\frac{\partial L}{\partial g_{\alpha\beta,\mu}} - \sum_{\nu=0}^{3}\frac{1}{n(\mu\nu)}D_\nu\left( \frac{\partial L}{\partial g_{\alpha\beta,\mu\nu}}\right)
\\
H&=&\sum_{\alpha\leq\beta}L^{\alpha\beta,\mu}g_{\alpha\beta,\mu}+\sum_{\alpha\leq\beta}L^{\alpha\beta,I}g_{\alpha\beta,I}-\sum_{\alpha\leq\beta}L \ .
\eeann
Finally, the corresponding Poincar\'e-Cartan $4$-form is
$\Omega_{\mathcal{L}_{EH}}=-\d\Theta_{\mathcal{L}_{EH}}$.

As $\Omega_{\mathcal{L}}$ is a premultisymplectic form, 
the field equations $\inn({\bf X})\Omega_{\Lag_{EH}}=0$ 
have no solution everywhere in $J^3\pi$,
but in a final constraint submanifold
${\cal S}\hookrightarrow J^3\pi$ which is locally defined by the
constraints \cite{GR1}
\beann
L^{\alpha\beta}:=
-\varrho\,n(\alpha\beta) (R^{\alpha\beta}-\frac{1}{2}g^{\alpha\beta}R)=0 \ .
\\
D_\tau L^{\alpha\beta}=
D_\tau(-\varrho\,n(\alpha\beta)(R^{\alpha\beta}-\frac{1}{2}g^{\alpha\beta}R))=0 \ .
\eeann
In particular,
\beann
{\bf{X}}_\mathcal{L}=&
\displaystyle\bigwedge_{\tau=0}^3 \sum_{\alpha\leq\beta}\sum_{\mu\leq\nu\leq\lambda}\Big(\derpar{}{x^\tau}+ g_{\alpha\beta,\tau}\frac{\partial}{\partial g_{\alpha\beta}}+ g_{\alpha\beta,\mu\tau}\frac{\partial}{\partial g_{\alpha\beta,\mu}}+
\\
&g_{\alpha\beta,\mu\nu\tau}\derpar{}{g_{\alpha\beta,\mu\nu}}+
D_\tau D_\lambda (g_{\lambda\sigma}(\Gamma_{\nu \alpha }^\lambda\Gamma_{\mu \beta}^\sigma+\Gamma_{\nu \beta}^\lambda\Gamma_{\mu \alpha }^\sigma))\derpar{}{g_{\alpha\beta,\mu\nu\lambda}}\Big) 
\eeann
is a holonomic multivector field solution to 
the equation in ${\cal S}$ and tangent to ${\cal S}$.
Then, their integral sections
$\psi(x)=(x^\mu,\,g_{\alpha\beta}(x),\,g_{\alpha\beta,\mu}(x),\,g_{\alpha\beta,\mu\nu}(x),
\,g_{\alpha\beta,\mu\nu\lambda}(x))$ 
 are the solutions to the equations 
\beann
g_{\alpha\beta,\mu}-\frac{\partial g_{\alpha\beta}}{\partial x^\mu}&=&0 
\qquad \mbox{\rm (holonomy conditions)} \ ,
\\
g_{\alpha\beta,\mu\nu}-\frac{1}{n(\mu\nu)}\Big(\frac{\partial g_{\alpha\beta,\mu}}{\partial x^{\nu}}+\frac{\partial g_{\alpha\beta,\nu}}{\partial x^{\mu}}\Big)&=&0 
\qquad \mbox{\rm (holonomy conditions)} \ ,
\\
\varrho\,n(\alpha\beta)(R^{\alpha\beta}-
\frac{1}{2}g^{\alpha\beta}R)&=&0 
\qquad \mbox{\rm (Einstein equations)} \ .
\eeann

Regarding the gauge vector fields,
 notice that,
as $\Omega_{\mathcal{L}_{EH}}$ is $\pi^3_1$-projectable \cite{first,Krupka,KrupkaStepanova,rosado,rosado2},
the $\pi^3_1$-vertical vector fields in $J^3\pi$ are gauge symmetries. It can be show that they are the only ones \cite{GR1}. In particular, there only exists one holonomic section in each gauge class.

We can analyze the Cartan or Noether symmetries for this system. 
First, we need to state some previous concepts 
\cite{Krupka,art:Munoz-rosado2013,art:Munoz-valdes1994}.
Remember that $\pi\colon E\to M$ is a bundle of metrics
and hence, if $p\equiv (x,g_x)\in E$, then $x\in M$ and
$g_x$ is a Lorentzian metric. Then:

\begin{definition}
\ben
\item
Let $F\colon M\to M$ be a diffeomorphism.
The \textbf{canonical lift of $F$ to the bundle of metrics $E$} 
is the diffeomorphism ${\cal F}\colon E\to E$ 
defined as follows: for every $(x,g_x)\in E$, then
${\cal F}(x,g_x):=(F(x),(F^{-1})^*(g_x))$.
(Thus $\pi\circ{\cal F}=F\circ\pi$).

The \textbf{canonical lift of ${\cal F}$ to the jet bundle
$J^k\pi$} is the diffeomorphism $j^k{\cal F}\colon J^k\pi\to J^k\pi$ 
defined as follows: for every $j^k_x\phi\in J^k\pi$, then
${\cal F}(j^k_x\phi):=j^k({\cal F}\circ\phi\circ F^{-1})(x)$.
\item
Let $Z\in\vf (M)$.
The \textbf{canonical lift of $Z$ to the bundle of metrics $E$}
is the vector field $Y\in\vf(E)$ whose associated
local one-parameter groups of diffeomorphisms  ${\cal F}_t$
are the canonical lifts to the bundle of metrics $E$
of the local one-parameter groups of diffeomorphisms $F_t$ of $Z$.

The \textbf{canonical lift of $Y\in\vf(E)$ to the jet bundle $J^k\pi$}
is the vector field $Y^k\equiv j^kY\in\vf(J^k\pi)$ whose associated
local one-parameter groups of diffeomorphisms are
the canonical lifts $j^1{\cal F}_t$ of the local
one-parameter groups of diffeomorphisms ${\cal F}_t$ of $Y$.
\een
\end{definition}

Observe that the canonical lifts $Y\in\vf(E)$ of
vector fields $Z\in\vf(M)$ 
to the bundle of metrics $E$ are $\pi$-projectable vector fields
and that $Y^k\in\vf(J^k\pi)$ are $\pi^k$ and $\bar\pi^k$ projectable vector fields.

In natural coordinates, if
$\displaystyle Z=u^\mu(x)\frac{\partial}{\partial x^\mu}\in\mathfrak{X}(M)$,
then the canonical lift of $Z$ to the bundle of metrics, $Y\in\mathfrak{X}(E)$, is given by
$$
Y=u^\mu\frac{\partial}{\partial x^\mu}-\sum_{\alpha\leq \beta}\left(\frac{\partial u^\mu}{\partial x^\alpha}g_{\mu\beta}+\frac{\partial u^\mu}{\partial x^\beta}g_{\mu\alpha}\right)\frac{\partial}{\partial g_{\alpha\beta}} \ ,
$$
and then we can lift this vector field $Y$ to the higher-order jet bundles $J^k\pi$; for instance,
\beann
Y^1&=& j^1Y=
u^\mu\frac{\partial}{\partial x^\mu}+\sum_{\alpha\leq \beta}Y_{\alpha\beta}\frac{\partial}{\partial g_{\alpha\beta}}+\sum_{\alpha\leq\beta}Y_{\alpha\beta\mu}\frac{\partial}{\partial g_{\alpha\beta,\mu}}
\\ &=&
u^\mu\frac{\partial}{\partial x^\mu}-\sum_{\alpha\leq \beta}\left(\frac{\partial u^\mu}{\partial x^\alpha}g_{\mu\beta}+\frac{\partial u^\mu}{\partial x^\beta}g_{\mu\alpha}\right)\frac{\partial}{\partial g_{\alpha\beta}}
\\ & &
-\sum_{\alpha\leq\beta}\left(\frac{\partial^2u^\nu}{\partial x^\alpha\partial x^\mu}g_{\nu\beta}+\frac{\partial^2u^\nu}{\partial x^\beta\partial x^\mu}g_{\alpha\nu}+\frac{\partial u^\nu}{\partial x^\alpha}g_{\nu\beta,\mu}+\frac{\partial u^\nu}{\partial x^\beta}g_{\alpha\nu,\mu}+\frac{\partial u^\nu}{\partial x^\mu}g_{\alpha\beta,\nu}\right)\frac{\partial}{\partial g_{\alpha\beta,\mu}} \ .
\eeann
Every $Z\in\mathfrak{X}(M)$ is an infinitessimal generator
of diffeomorphisms in $M$. Then, if $Y^3=j^3Y$, we have that 
$\Lie (Y^3)\mathcal{L}_{EH}=0$, because $\mathcal{L}_{EH}$ is invariant under diffeomorphisms.
Furthermore, as $Y^3$ is a canonical lift, 
it is an infinitesimal Lagrangian symmetry and thus, 
by Proposition \ref{prop:jet}, $Y^3$ it is an exact infinitesimal Cartan symmetry. 
The conserved quantity associated to $Y^3$ is 
$\xi_Y=\inn(Y^3)\Theta_{\mathcal{L}_{EH}}$ and,
as $\Theta_{\Lag_{EH}}$ is a $\pi^3_1$-basic form, 
we have that
\beann
\xi_Y&=&\inn(Y^3)\Theta_{\mathcal{L}_{EH}}=\inn(Y^1)\Theta_{\mathcal{L}_{EH}} 
=\left(u^\mu H + Y_{\alpha\beta}L^{\alpha\beta,\mu}+
Y_{\alpha\beta\nu}L^{\alpha\beta,\nu\mu}\right)\d^3x_\mu
\\ & &
+\left(u^\nu L^{\alpha\beta,\mu}-u^\mu L^{\alpha\beta,\nu}\right)\d g_{\alpha\beta}\wedge\d^2x_{\mu\nu}
+\left(u^\nu L^{\alpha\beta,\lambda\mu}-u^\mu L^{\alpha\beta,\lambda\nu}\right)\d g_{\alpha\beta,\lambda}\wedge\d^2x_{\mu\nu} \ ,
\eeann
where $\displaystyle\d^2 x_{\mu\nu}=
\inn\left(\frac{\partial}{\partial x^\nu}\right)\inn\left(\frac{\partial}{\partial x^\mu}\right)\d^4x$.
The vector fields of the form $Y$ are the only 
natural infinitesimal Lagrangian symmetries 
\cite{Krupka,rosado2}.

\subsection{The Einstein-Palatini action (metric-affine model)}

Now we consider the Einstein-Palatini (or metric-affine) action for the Einstein equations without sources.

The configuration bundle for this system is $\pi\colon E\rightarrow M$, 
where $M$ is a connected 4-dimensional manifold representing space-time and
$E$ is $\Sigma\times_MC(LM)$, where $\Sigma$ is the manifold of Lorentzian metrics on $M$ and $C(LM)$ is the bundle of connections on $M$; that is, linear connections in $\Tan M$.
We use the local coordinates in $E$ $(x^\mu,g_{\alpha\beta},\Gamma^\alpha_{\beta\gamma})$,
 with $0\leq\alpha\leq\beta\leq3$.
 We \emph{do not} assume torsionless connections; thus, in general, $\Gamma^\alpha_{\beta\gamma}\neq\Gamma^\alpha_{\gamma\beta}$.
The induced coordinates in $J^1\pi$ are 
$(x^\mu,\,g_{\alpha\beta},\,g_{\alpha\beta,\mu},\,\Gamma^\alpha_{\beta\gamma},
\,\Gamma^\alpha_{\beta\gamma,\mu})$.

Using this coordinates, the local expression of the Palatini-Einstein Lagrangian is
$$
L_{PE}=\sqrt{|{\rm det}(g)|}\,g^{\alpha\beta}R_{\alpha\beta}\equiv
\varrho g^{\alpha\beta}R_{\alpha\beta}
$$
where, as above, $\varrho\equiv\sqrt{|det(g_{\alpha\beta})|}$, $R_{\alpha\beta}=\Gamma^{\gamma}_{\beta\alpha,\gamma}-\Gamma^{\gamma}_{\gamma\alpha,\beta}+
\Gamma^{\gamma}_{\beta\alpha}\Gamma^{\sigma}_{\sigma\gamma}-
\Gamma^{\gamma}_{\beta\sigma}\Gamma^{\sigma}_{\gamma\alpha}$
are the components of the {\sl Ricci tensor}, which 
now depend only on the connection.
We consider the auxiliary functions
\bea \nonumber
L^{\beta\gamma,\mu}_{\alpha}:=\frac{\partial L_{PE}}{\partial \Gamma^{\alpha}_{\beta\gamma,\mu}}=\varrho\frac12(\delta_\alpha^\mu g^{\beta\gamma}-\delta_\alpha^\beta g^{\mu\gamma})\,,
\\\nonumber
H:=L^{\beta\gamma,\mu}_{\alpha}\Gamma^{\alpha}_{\beta\gamma,\mu}-L_{PE}=\varrho g^{\alpha\beta}\left(\Gamma^{\gamma}_{\beta\sigma}\Gamma^{\sigma}_{\gamma\alpha}-\Gamma^{\gamma}_{\beta\alpha}\Gamma^{\sigma}_{\sigma\gamma}\right)\ .
\eea
Then, the Poincar\'e-Cartan $4$-form $\Omega_{\mathcal{L}_{EP}}$ 
associated with the Einstein-Palatini Lagrangian density
$\mathcal{L}_{EP}=
L_{EP}\,(\overline{\pi}^2)^*\omega=L_{Ep}\,\d^4x$ is
\beann
&\Omega_{\mathcal{L}_{EP}}&=\d H\wedge\d^4x -\d L^{\beta\gamma,\mu}_{\alpha}\wedge\d \Gamma^{\alpha}_{\beta\gamma}\wedge \d^{m-1}x_{\mu}  \ ;
\eeann

As above, the field equations $\inn({\bf X})\Omega_{\Lag_{EP}}=0$ 
have no solution everywhere in $J^3\pi$.
Now, the premultisymplectic constraint algorithm leads to the constraints (see  \cite{GR2}):
\beann
0&=&\frac{\partial H}{\partial g_{\mu\nu}}-
\frac{\partial L_\alpha^{\beta\gamma,\sigma}}{\partial g_{\mu\nu}}\Gamma^\alpha_{\beta\gamma,\sigma} \equiv c^{\mu\nu} \ ,
\\
0&=&g_{\rho\sigma,\mu}-g_{\sigma\lambda}\Gamma^\lambda_{\mu\rho}-
g_{\rho\lambda}\Gamma^\lambda_{\mu\sigma}-\frac{2}{3}g_{\rho\sigma}T^\lambda_{\lambda\mu} \equiv m_{\rho\sigma,\mu} \ ,
\\
0&=&T^\alpha_{\beta\gamma}-
\frac13\delta^\alpha_\beta T^\mu_{\mu\gamma}+
\frac13\delta^\alpha_\gamma T^\mu_{\mu\beta} \equiv t^\alpha_{\beta\gamma} \ ,
\\
0&=&T^\alpha_{\beta\gamma,\nu}-
\frac13\delta^\alpha_\beta T^\mu_{\mu\gamma,\nu}+
\frac13\delta^\alpha_\gamma T^\mu_{\mu\beta,\nu} \equiv  r^\alpha_{\beta\gamma,\nu}\ ,
\\
0&=&g_{\rho\gamma}\Gamma^\gamma_{[\nu\lambda}\Gamma^\lambda_{\mu]\sigma}+g_{\sigma\gamma}\Gamma^\gamma_{[\nu\lambda}\Gamma^\lambda_{\mu]\rho}+g_{\rho\lambda}\Gamma^\lambda_{[\mu\sigma,\nu]}+g_{\sigma\lambda}\Gamma^\lambda_{[\mu\rho,\nu]}+\frac23g_{\rho\sigma}T^\lambda_{\lambda[\mu,\nu]} \equiv i_{\rho\sigma,\mu\nu}\ . 
\eeann
where 
$T^\alpha_{\beta\gamma}\equiv\Gamma^\alpha_{\beta\gamma}-\Gamma^\alpha_{\gamma\beta}$.
They define the submanifold $\jmath_{\cal S}\colon{\mathcal S}\hookrightarrow J^1\pi$,
where there are holonomic multivector fields solution to
the field equations in ${\cal S}$ and tangent to ${\mathcal S}$.
The holonomic multivector fields
which are solutions to the field equations on ${\cal S}$ are
$$
\mathbf{X}=
\bigwedge_{\nu=0}^3\left(\frac{\partial}{\partial x^\nu}+
\sum_{\rho\leq\sigma}\left(g_{\rho\sigma,\nu}\frac{\partial}{\partial g_{\rho\sigma}}+
f_{\rho\sigma\mu,\nu}\frac{\partial}{\partial g_{\rho\sigma,\mu}}\right)+
\Gamma^\alpha_{\beta\gamma,\nu}\frac{\partial}{\partial \Gamma^\alpha_{\beta\gamma}}+
f^\alpha_{\beta\gamma\mu,\nu}\frac{\partial}{\partial \Gamma^\alpha_{\beta\gamma,\mu}}
\right) \ .
$$

As the Einstein-Palatini action only has solutions on ${\cal S}$,
to get the gauge fields of the theory is more laborious. 
After some computation it can be shown that 
$$
\ker^{V(\bar\pi^1)}\Omega_\Lag=
{\left<\delta^\alpha_\gamma\frac{\partial}{\partial \Gamma^\alpha_{\beta\gamma}},K^\alpha_{\beta\gamma}\frac{\partial}{\partial \Gamma^\alpha_{\beta\gamma}},  
\frac{\partial}{\partial g_{\alpha\beta,\mu}},\frac{\partial}{\partial \Gamma^\alpha_{\beta\gamma,\mu}}\right>} \ ,
$$
where $K^\nu_{\nu\gamma}=0$ and
$K^{\nu}_{\beta\gamma }+K^\nu_{\gamma\beta }=0$.
The  vector fields $\displaystyle K^\alpha_{\beta\gamma}\frac{\partial}{\partial \Gamma^\alpha_{\beta\gamma}}$ are not tangent to $\mathcal{S}$, thus they are not gauge vector fields. 
The vectors field of the form $\displaystyle\delta^\alpha_\gamma\frac{\partial}{\partial \Gamma^\alpha_{\beta\gamma}}$ are tangent to $\mathcal{S}$
and therefore they are gauge vector fields. 
The other vector fields, 
$\displaystyle \frac{\partial}{\partial g_{\alpha\beta,\mu}},\ \frac{\partial}{\partial \Gamma^\alpha_{\beta\gamma,\mu}}$,
arise from the projectability of the theory. 
Only one representative is holonomic and consistent with a gauge fixing for 
$\displaystyle\delta^\alpha_\gamma\frac{\partial}{\partial \Gamma^\alpha_{\beta\gamma}}$.

Finally, it is relevant to point out that a gauge fixing
in the Einstein-Palatini model leads to recover the Einstein-Hilbert model (see \cite{pons,GR2}).

Concerning to Noether symmetries,
the Einstein-Palatini Lagrangian $\mathcal{L}_{\rm EP}$ 
is invariant under diffeomorphisms in $M$
(as it can be checked using the constraints $c^{\mu\nu}$).
Then we have to consider the canonical lift of vector fields 
$\displaystyle Z=f^\mu(x)\frac{\partial}{\partial x^\mu}\in\mathfrak{X}(M)$
to the bundle $E\to M$, which is now
\beann
Y_Z&=&f^\mu\frac{\partial}{\partial x^\mu}-\sum_{\alpha\leq \beta}\left(\frac{\partial f^\lambda}{\partial x^\alpha}g_{\lambda\beta}+\frac{\partial f^\lambda}{\partial x^\beta}g_{\lambda\alpha}\right)\frac{\partial}{\partial g_{\alpha\beta}}
\\ & &
+\left(\frac{\partial f^\alpha}{\partial x^\lambda}\Gamma^\lambda_{\beta\gamma}-\frac{\partial f^\lambda}{\partial x^\beta}\Gamma^\alpha_{\lambda\gamma}-\frac{\partial f^\lambda}{\partial x^\gamma}\Gamma^\alpha_{\beta\lambda}-\frac{\partial^2f^\alpha}{\partial x^\beta\partial x^\gamma}\right)\frac{\partial}{\partial \Gamma^\alpha_{\beta\gamma}}
\in\vf({\rm E})\  .
\eeann
and then
\beann
j^1Y_Z&=&
f^\mu\frac{\partial}{\partial x^\mu}-\sum_{\alpha\leq \beta}\left(\frac{\partial f^\lambda}{\partial x^\alpha}g_{\lambda\beta}+\frac{\partial f^\lambda}{\partial x^\beta}g_{\lambda\alpha}\right)\frac{\partial}{\partial g_{\alpha\beta}}-
\\ & &
\sum_{\alpha\leq\beta}\left(\frac{\partial^2f^\nu}{\partial x^\alpha\partial x^\mu}g_{\nu\beta}+\frac{\partial^2f^\nu}{\partial x^\beta\partial x^\mu}g_{\alpha\nu}+\frac{\partial f^\nu}{\partial x^\alpha}g_{\nu\beta,\mu}+\frac{\partial f^\nu}{\partial x^\beta}g_{\alpha\nu,\mu}+\frac{\partial f^\nu}{\partial x^\mu}g_{\alpha\beta,\nu}\right)\frac{\partial}{\partial g_{\alpha\beta,\mu}}+
\\ & &
\left(\frac{\partial f^\alpha}{\partial x^\lambda}\Gamma^\lambda_{\beta\gamma}-\frac{\partial f^\lambda}{\partial x^\beta}\Gamma^\alpha_{\lambda\gamma}-\frac{\partial f^\lambda}{\partial x^\gamma}\Gamma^\alpha_{\beta\lambda}-\frac{\partial^2f^\alpha}{\partial x^\beta\partial x^\gamma}\right)\frac{\partial}{\partial \Gamma^\alpha_{\beta\gamma}}+
\\ & &
\left(\frac{\partial f^\alpha}{\partial x^\lambda}\Gamma^\lambda_{\beta\gamma,\mu}-\frac{\partial f^\lambda}{\partial x^\beta}\Gamma^\alpha_{\lambda\gamma,\mu}-\frac{\partial f^\lambda}{\partial x^\gamma}\Gamma^\alpha_{\beta\lambda,\mu}-\frac{\partial f^\lambda}{\partial x^\mu}\Gamma^\alpha_{\beta\gamma,\lambda}+\right.
\\& &
\left.\frac{\partial^2 f^\alpha}{\partial x^\lambda\partial x^\mu}\Gamma^\lambda_{\beta\gamma}-\frac{\partial^2 f^\lambda}{\partial x^\beta\partial x^\mu}\Gamma^\alpha_{\lambda\gamma}-\frac{\partial^2 f^\lambda}{\partial x^\gamma\partial x^\mu}\Gamma^\alpha_{\beta\lambda}-
\frac{\partial^3f^\alpha}{\partial x^\beta\partial x^\gamma\partial x^\mu}\right)\frac{\partial}{\partial \Gamma^\alpha_{\beta\gamma,\mu}}
\\ &\equiv&
f^\mu\frac{\partial}{\partial x^\mu}+\sum_{\alpha\leq \beta}Y_{\alpha\beta}\frac{\partial}{\partial g_{\alpha\beta}}+\sum_{\alpha\leq\beta}Y_{\alpha\beta\mu}\frac{\partial}{\partial g_{\alpha\beta,\mu}}
+ Y^\alpha_{\beta\gamma}\frac{\partial}{\partial\Gamma^\alpha_{\beta\gamma}}+Y^\alpha_{\beta\gamma\mu}\frac{\partial}{\partial \Gamma^\alpha_{\beta\gamma,\mu}}
\in\vf(J^1\pi)\ .
\eeann
As a long calculation shows,
$j^1Y_Z$ are tangent to ${\cal S}$ since
\beann
\Lie(j^1Y_Z)c^{\mu\nu}=\left(-\frac{\partial f^\mu}{\partial x^\rho}\delta^\nu_{\sigma}-\frac{\partial f^\nu}{\partial x^\sigma}\delta^\mu_{\rho}\right)\left(\frac{\partial H}{\partial g_{\rho\sigma}}-\frac{\partial L^{\beta\gamma,\lambda}_\alpha}{\partial g_{\rho\sigma}}\Gamma^{\alpha}_{\beta\gamma,\lambda}\right)=0; \ \mbox{\rm (on ${\cal S}$)}\, ,
\\
\Lie(j^1Y_Z)m_{\rho\sigma,\mu}= 
\left(-\frac{\partial f^\alpha}{\partial x^\rho}\delta^\beta_{\sigma}\delta^\nu_{\mu}-\frac{\partial f^\beta}{\partial x^\sigma}\delta^\alpha_{\rho}\delta^\nu_{\mu}-\frac{\partial f^\nu}{\partial x^\mu}\delta^\alpha_{\rho}\delta^\beta_{\sigma}\right)m_{\alpha\beta,\nu} =0;
\ \mbox{\rm (on ${\cal S}$)} \, ,
\\
\Lie(j^1Y_Z)t^\alpha_{\beta\gamma}= \left(\frac{\partial f^\alpha}{\partial x^\lambda}\delta^\rho_{\beta}\delta^\sigma_{\gamma}-\frac{\partial f^\rho}{\partial x^\beta}\delta^\alpha_{\lambda}\delta^\sigma_{\gamma}-\frac{\partial f^\sigma}{\partial x^\gamma}\delta^\alpha_{\lambda}\delta^\rho_{\beta}\right)t^\lambda_{\rho\sigma}=0;
\ \mbox{\rm (on ${\cal S}$)} \, ,
\\
\Lie(j^1Y_Z)r^\alpha_{\beta\gamma,\nu}= \left(\frac{\partial f^\alpha}{\partial x^\lambda}\delta^\rho_{\beta}\delta^\sigma_{\gamma}\delta^\tau_{\nu}-\frac{\partial f^\rho}{\partial x^\beta}\delta^\alpha_{\lambda}\delta^\sigma_{\gamma}\delta^\tau_{\nu}-\frac{\partial f^\sigma}{\partial x^\gamma}\delta^\alpha_{\lambda}\delta^\rho_{\beta}\delta^\tau_{\nu}-\frac{\partial f^\tau}{\partial x^\nu}\delta^\alpha_{\lambda}\delta^\rho_{\beta}\delta^\sigma_{\gamma}\right)r^\lambda_{\rho\sigma,\tau}=0;
\ \mbox{\rm (on ${\cal S}$)} \, ,
\\
\Lie(j^1Y_Z)i_{\rho\sigma,\mu\nu}= 
\left(-\frac{\partial f^\alpha}{\partial x^\rho}\delta^\beta_{\sigma}\delta^\lambda_{\mu}\delta^\gamma_\nu-\frac{\partial f^\beta}{\partial x^\sigma}\delta^\alpha_{\rho}\delta^\lambda_{\mu}\delta^\gamma_\nu-\frac{\partial f^\lambda}{\partial x^\mu}\delta^\alpha_{\rho}\delta^\beta_{\sigma}\delta^\gamma_\nu-\frac{\partial f^\gamma}{\partial x^\nu}\delta^\alpha_{\rho}\delta^\beta_{\sigma}\delta^\lambda_\mu\right)i_{\alpha\beta,\lambda\gamma} =0;
\ \mbox{\rm (on ${\cal S}$)} \, .
\eeann
Furthermore,
$\Lie (j^1Y_Z)\mathcal{L}_{\rm EP}\vert_{\mathcal S}=0$,
for every $Z\in\mathfrak{X}(M)$, and
as these are natural vector fields, 
the Euler-Lagrange equations are also invariant.
Thus, they are natural infinitesimal Lagrangian symmetries 
and, hence, natural infinitesimal Noether symmetries. 
The associated conserved quantity to each $j^1Y_Z$ is
$$
\xi_{Y_Z}=\inn(j^1Y)\Theta_{\mathcal{L}_{\rm EP}}=
(L^{\beta\gamma,\mu}_\alpha Y^\alpha_{\beta\gamma}-Hf^\mu)\d^3x_\mu+f^\mu L_\alpha^{\beta\gamma,\nu}\d \Gamma^\alpha_{\beta\gamma}\wedge\d^2 x_{\mu\nu} \ ;
$$
and, given a section $\psi$ solution to the field equations, 
the current associated with $j^1Y_Z$ is
$$
\psi^*\xi_{Y_Z}=
\psi^*(L^{\beta\gamma,\mu}_\alpha(Y^\alpha_{\beta\gamma}-\Gamma^\alpha_{\beta\gamma,\lambda}f^\lambda)-f^\mu L_{\rm EP})\d^3x_\mu \ .
$$


\section{Conclusions and outlook}
\protect\label{di}

In this work a careful review has been made on the geometric meaning of symmetries
for Lagrangian field theories.
The discussion has been done for first and second-order Lagrangians 
because our aim was to apply the concepts and results
to different cases of classical field theories;
namely, the Maxwell theory of Electromagnetism (first-order)
and the Einstein--Hilbert (second-order) and the Einstein--Palatini (first-order) 
models in General Relativity.

First, we have stated the geometrical meaning of {\sl symmetry}
in the ambient of the multisymplectic framework of field theories.
Then, we have introduced the so-called {\sl Cartan} or {\sl Noether symmetries},
which are associated with {\sl conservation laws} by means of {Noether's theorem}. 
These kinds of symmetries are geometrically characterized by the fact that they let the
multisymplectic {\sl Poincar\'e--Cartan $(n+1)$-form} invariant,
and this property allows us to obtain {\sl conserved quantities}
as $(n-1)$-forms satisfying certain properties which have been carefully analyzed
and that lead to state the corresponding {\sl conservation laws}.
(Most of these are well-known results, previously studied in \cite{art:deLeon_etal2004,EMR-99b,art:GPR-2016,RWZ-2016},
as well as in other works cited therein).

We have also studied {\sl gauge symmetries} from a pure geometric perspective.
These symmetries generate transformations that have no physical relevance and 
are associated to singular Lagrangians and, hence, to
premultisymplectic Poincar\'e--Cartan forms.
The definition and characteristic properties of vector fields
generating {\sl gauge transformations} have been discussed and
justified, and
this leads to take the vertical vector fields in the kernel of the Poincar\'e--Cartan form as
infinitesimal generators of these symmetries.
The concept of {\sl gauge equivalent solutions} to the field equations
and of {\sl gauge equivalent states} has been also discussed.
Finally, we have explained the guidelines of {\sl gauge reduction}
which allows us to eliminate the non-physical degrees of freedom of the theory associated with gauge symmetries.
All these ideas are the generalization to premultisymplectic field theories
of the analysis made on this topic for presymplectic mechanical systems, mainly in \cite{BK-86,GN-79}.

As an application of the above ideas,
we have analyzed the symmetries, conservation laws and gauge content
of the aforementioned cases:
Electromagnetism, and the Einstein--Hilbert and the Einstein--Palatini models of Gravitation;
recovering the already well-known results of these theories
(see, for instance, \cite{GR1,GR2,GIMMSY,Krupka,rosado2}).

Following this guidelines,
all this methods could be applied to investigate
Noether symmetries and conservation laws,
as well as gauge symmetries for other theories of gravity (for instance, {\sl Chern-Simons gravity\/})
and other extended models of  General Relativity \cite{CdL-2011,art:Capriotti3,CGM-16,Fe-12}.

The geometric interpretation of gauge symmetries in the multisymplectic context is clearly incomplete. 
Looking at the case of Electromagnetism, one could argue that the gauge freedom in the physical sense is better understood as a Lagrangian symmetry than as a gauge symmetry. 
This points out that it would be relevant to study the interplay between this two kind of symmetries, and the role of the holonomy condition. 
Moreover, the classical works \cite{BK-86,GN-79} for mechanical systems show a more complex structure of gauge vector fields and a more consistent reduction. It would be interesting to generalize these results to the multisymplectic case, if possible.

Finally, there is the important problem on how to use conserved currents to integrate the field equations. 
We expect that the complete characterization of all the types of symmetries 
of multisymplectic systems could be relevant for this problem.


\section*{Acknowledgments}

We acknowledge the financial support of the 
{\sl Ministerio de Ciencia, Innovaci\'on y Universidades} (Spain), project 
PGC2018-098265-B-C33, and of
{\sl Generalitat de Catalunya}, project 2017–SGR–932.


\end{document}